\documentclass[pdftex,twocolumn,epjc3]{svjour3}          

\RequirePackage[T1]{fontenc}

\smartqed  

\RequirePackage{graphicx}
\RequirePackage{mathptmx}      
\RequirePackage{flushend}
\RequirePackage[numbers,sort&compress]{natbib}
\RequirePackage[colorlinks,citecolor=blue,urlcolor=blue,linkcolor=blue]{hyperref}
\usepackage{subfigure}

\usepackage{microtype}
\usepackage{multirow}
\usepackage{comment}
\journalname{Eur. Phys. J. C}

\begin{document}

\title{Simulation of a radial TPC for the detection of neutrinoless double beta decay}

\author{R.~Bouet \thanksref{LP2I}  
	\and
	J.~Busto \thanksref{CPPM}
          \and
         A.~Cadiou \thanksref{SUBATECH}  
         \and
        P.~Charpentier \thanksref{LP2I}  
         \and
         D.~Charrier \thanksref{SUBATECH}  
         \and
        M.~Chapellier \thanksref{IJCLab}  
        \and
        A.~Dastgheibi-Fard \thanksref{LPSC}  
           \and
        F.~Druillole \thanksref{LP2I}  
           \and
        P.~Hellmuth \thanksref{LP2I} 
            \and
        C.~Jollet \thanksref{LP2I}  
            \and
        J.~Kaizer \thanksref{Bratislava} 
            \and
         I.~Kontul \thanksref{Bratislava} 
            \and
        P.~Le Ray \thanksref{SUBATECH}  
            \and
        M.~Gros \thanksref{CEA}  
            \and
        P.~Lautridou \thanksref{e2,SUBATECH}  
            \and
        M.~Macko \thanksref{Prague}  
                    \and
        A.~Meregaglia \thanksref{e3,LP2I}  
            \and
        F.~Piquemal \thanksref{LP2I}  
       \and
        P.~Povinec \thanksref{Bratislava}  
       \and
        M.~Roche \thanksref{LP2I}  
    	}

\thankstext{e3}{e-mail: anselmo.meregaglia@cern.ch}
\thankstext{e2}{e-mail: pascal.lautridou@subatech.in2p3.fr}

\institute{LP2I Bordeaux, Universit\'e de Bordeaux, CNRS/IN2P3, F-33175 Gradignan, France\label{LP2I}
\and	
CPPM, Universit\'e d'Aix-Marseille, CNRS/IN2P3, F-13288 Marseille, France\label{CPPM}
\and
SUBATECH, IMT-Atlantique, Universit\'e de Nantes, CNRS-IN2P3, France\label{SUBATECH}
\and
IJCLab, CNRS/IN2P3, Paris, France\label{IJCLab}
\and
LPSC-LSM, CNRS/IN2P3, Universit\'e Grenoble-Alpes, Modane, France\label{LPSC}
\and
IRFU, CEA, Universit\'e Paris-Saclay, F-91191 Gif-sur-Yvette, France\label{CEA}
\and
IEAP, Czech Technical University in Prague, CZ-11000 Prague, Czech Republic
\label{Prague}
\and
Faculty of Mathematics, Physics and Informatics, Comenius University, Bratislava, Slovakia\label{Bratislava}
}

\date{Received: date / Accepted: date}

\maketitle

\begin{abstract}
To search for $\beta\beta0\nu$ decay with unprecedented sensitivity, the R2D2 collaboration is developing a radial time projection chamber with a fiducial mass of half a tonne of $^{136}$Xe at high pressure. The various approaches implemented to eliminate the radioactive background are presented in terms of detector design, topological recognition of interactions, and event energy reconstruction. The developed tools enable the disentangling of the sought-after signal from the background. The projected sensitivity after ten years of data taking yields a half-life limit exceeding $10^{27}$ years, along with a constraint on the effective neutrino mass $m_{\beta\beta}$ that could cover a large fraction of the inverted mass hierarchy region, depending on the final experimental background.
\end{abstract}

\section{Introduction}

The search for neutrinoless double beta decay ($\beta\beta0\nu$) is essential for determining whether neutrinos are Dirac or Majorana particles, {\it i.e.} whether they are distinct from their antiparticles or identical to them. Identifying neutrinos as Majorana particles would imply lepton number violation, a critical feature of leptogenesis models that explain the matter-antimatter asymmetry in the universe~\cite{Bodeker:2020ghk,Agostini:2022zub}. $\beta\beta0\nu$ is the only process that can confirm the Majorana nature of neutrinos, as it occurs only if the neutrino is its own antiparticle~\cite{DellOro:2016tmg}. Additionally, detecting $\beta\beta0\nu$ would provide insights into the absolute neutrino mass, an unresolved property of these particles. Thus, the search for this rare decay is pivotal for advancing both particle physics and cosmology.

Given the importance of such a potential discovery, the R2D2 (Rare Decays with Radial Detector) R\&D program, initiated in 2017, has been developing a high-pressure, single-anode, radial Time Projection Chamber (TPC) specifically aimed at detecting $\beta\beta0\nu$.

Experimentally, three key factors play a fundamental role in detector design: low background, large mass, and good energy resolution. 
Several ton-scale projects, such as LEGEND~\cite{LEGEND:2022bzq}, CUORE~\cite{CUORE:2022uex}, and nEXO~\cite{nEXO:2017nam}, aim to achieve an effective Majorana neutrino mass, $m_{\beta\beta}$, of $\sim 10$~meV, which would rule out the inverted hierarchy region in the absence of any signals.
Each technology has its own advantages but also presents some weaknesses. The R2D2 guideline has been to strike a balanced compromise, meeting all the critical requirements for the $\beta\beta0\nu$ search simultaneously.

The energy resolution of a spherical TPC (SPC for Spherical Proportional Chamber) was the primary goal of the R2D2 R\&D~\cite{Bouet:2020lbp,Bouet:2023zyk,Laut:2023vrk,Bouet:2023jpm}. A cylindrical geometry (CPC for Cylindrical Proportional Chamber) was also studied and found to be an optimal solution when operated in the ionization regime~\cite{Bouet:2024njk}. It was demonstrated that a resolution of $\sim$1.2\% Full Width at Half Maximum (FWHM) could be achieved up to 10~bar with point-like and diffuse $\alpha$ sources of $\sim 5.5$~MeV, regardless of the gas used (argon or xenon).
\begin{figure*}[h!]
    \centering
    \includegraphics[width=0.8\textwidth]{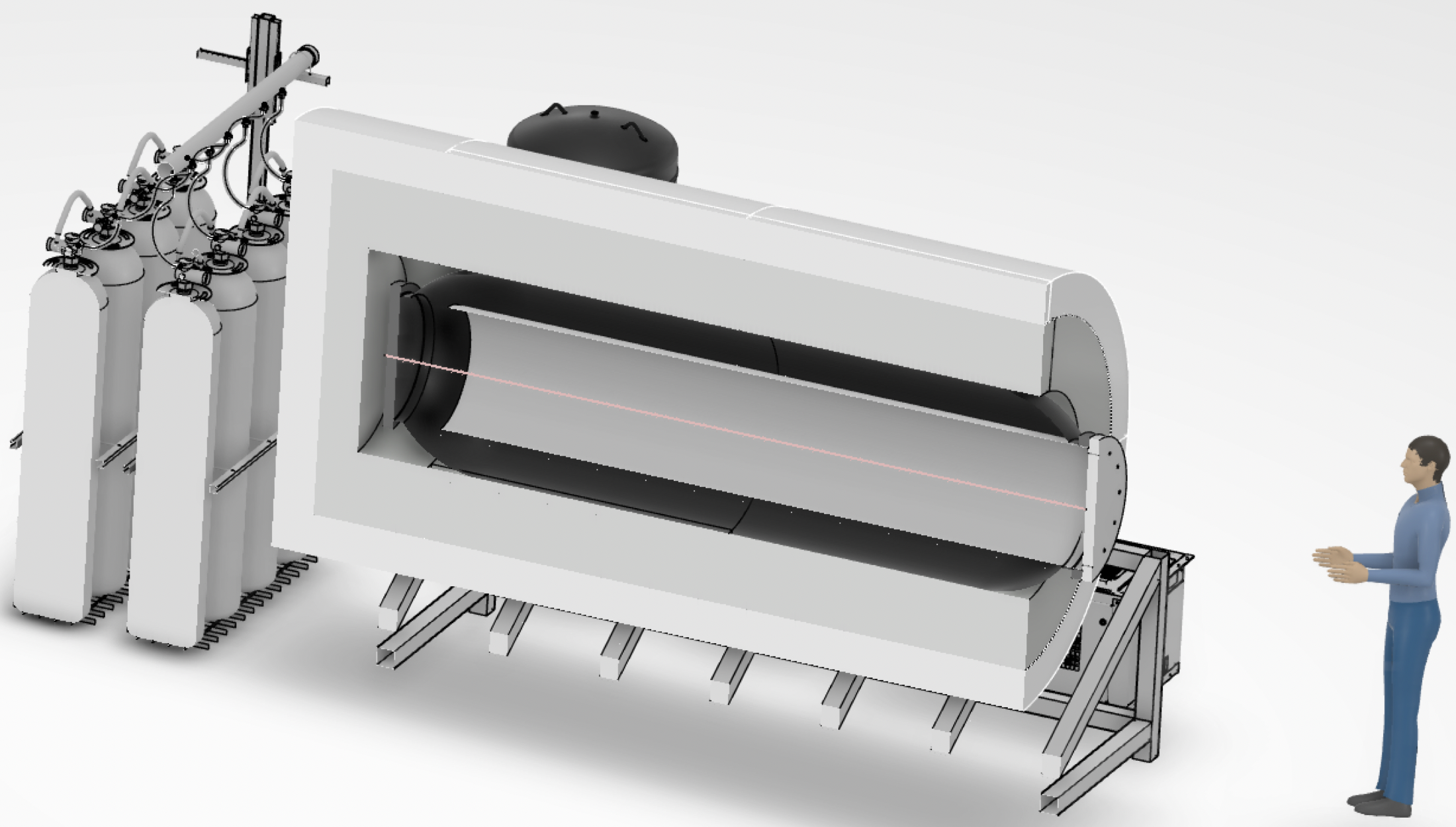}
    \caption{Conceptual illustration of the detector setup.}
    \label{fig:Detector}
\end{figure*}
In parallel, the issue of low background was addressed by reducing the total material budget of the detector, a central consideration in the design process, which utilizes a single readout channel.
A significant advancement could be achieved through the use of a thin composite vessel to contain the pressurized gas. Ongoing studies on different carbon fibers and glues used to construct this gas tank are being conducted in collaboration with industry partners, (IRT Jules Verne~\cite{IRT}).

The cylindrical active volume, which is 2.5~m long with a diameter of 1~m, can accommodate an active xenon mass of 582~kg at 40 bars. Enclosed in a tank, this simple detection system could therefore be competitive with current ton-scale projects. The cost of the proposed detector should be much lower than that of currently identified experiments and would not require cryogenic infrastructure.
Preliminary sensitivity studies on a spherical TPC were published in 2018~\cite{Meregaglia:2017nhx}, demonstrating that, in principle, a background-free detector at the ton scale was feasible.

In the present work, a comprehensive sensitivity study of a ton-scale CPC is presented, based on the R\&D results obtained over the last six years. A sensitivity on the half-life $T^{0\nu}_{1/2}$ greater than $10^{27}$ years can be achieved after ten years of data taking, corresponding to an effective neutrino mass $m_{\beta\beta}$ in the range of  13 to 57~meV. Such a limit can be reduced to 8--35~meV if the ultimate R2D2 goal of reaching zero background is achieved.

\section{Detector Setup}
\label{Sec:Det}

The detector geometry selected for the current sensitivity study is shown in the schematic view presented in Fig.~\ref{fig:Detector}. The main components are described below. The mechanical structure inherently minimizes the material budget, ensuring the lowest possible instrumental radioactivity. Although some elements, such as gas recirculation and purification, are not directly related to the detector simulation, all components are addressed for completeness..

\begin{itemize}
\item {\it Central anode}\\ 
The detector operates in ionization mode ({\it i.e.} without avalanche at the anode). A detailed study on the anode is ongoing to minimise its radioactivity. The most promising option, currently under testing and considered in the following simulation, is a hollow tube with a thickness of 1~mm and an external radius of 1~cm, made of polymer materials coated with a resistive foil a few micrometres thick.
The choice of radius is based on tests conducted on a previous CPC prototype, which demonstrated that a thick anode tube creates a higher drift field away from the anode, mitigating charge loss effects, and reduces geometrical distortions due to anode inhomogeneities. Improved energy resolution was indeed achieved with thick anodes in ionization mode~\cite{Bouet:2024njk,FabriceTPC}. The anode is grounded and connected to the readout electronics at both ends.
Grounding the anode and applying a negative high voltage to the cathode decouples the signal readout from the high voltage, thereby reducing electronic noise on the signal.

\item {\it Xenon volume}\\
  The tank, designed to withstand a maximum gas pressure of 60 bars, has a radius of 50~cm and a height of 350~cm, corresponding to a volume of 2.49~m$^3$. This results in a xenon mass of 740~kg . 
 By excluding the two rounded ends of the tank where the field is distorted (and where the anode is electrically masked), the active volume is reduced to the straight cylindrical section, which is 2.5 ~m high with diameter of 1~m. This corresponds to a useful volume of 1.96~m$^3$ and a $^{136}$Xe mass of 582~kg at 40~bars.
 
\item {\it Xenon vessel}\\
The xenon vessel design is based on the use of composite materials. These materials consist of a mix of carbon fibers and epoxy glue. Composite tanks, few millimeter thick and withstanding pressures of several hundred bars, are already used today in space industry for rocket fuel tanks and in the automotive industry for hydrogen reservoirs.
The structure used in the present sensitivity study is made of  1.5~cm of carbon fiber-based composite material. This is very similar to the tank currently under test, produced by MAHYTEC company~\cite{MAHYTEC}.\\
Dedicated studies are ongoing in cooperation with IRT Jules Verne company~\cite{IRT} to reduce the thickness and select low radioactivity materials. Several measurements of epoxy glues for low-background experiments have already been conducted~\cite{Busto:2002hq}, and some measurements of carbon fibres were also carried out, for example, by the nEXO collaboration~\cite{nEXO:2018ylp}. While carbon fibres are inherently low in contaminants, they can become contaminated with uranium and thorium during the industrial sizing process, which prepares them for integration with the glue. Current measurements of carbon fibers and resins are being conducted in partnership with industry, aiming for a composite material with a radioactivity level of 10~$\mu$Bq/kg.\\
A thin layer (200~$\mu$m) of aluminized mylar or another conducting material is deployed as a cathode on the inner side of the tank. This layer is polarized to a high voltage of -20~kV.

\item {\it Shielding}\\
To shield the active volume from external radioactivity, a combination of lead and polyethylene surrounds the tank. Starting from the inside, a layer of 35~cm of lead is used primarily to block external gamma rays, accounting for a total mass of about 75~ton. An outer layer of 30~cm of polyethylene, designed to stop external neutrons, encloses the previous shell.

\item {\it Electronics}\\
In the previous R\&D prototype, the anodic signal was read only on one side by a charge amplifier: a custom-made amplifier for the proportional mode~\cite{Bouet:2020lbp}, and an ORTEC one for the ionization mode~ \cite{Bouet:2024njk}. The waveforms were sampled at 2~MS/s by a CALI card (ADC), controlled by the SAMBA acquisition software~\cite{EDELWEISS:2017lvq}. 
In the present development, in order to achieve longitudinal localization, both sides of the anode must be grounded via specific amplifiers. These new devices (transimpedance  ASIC amplifiers) are being developed based on SiGe technology. For the present simulation, to ensure our interpretation (in terms of noise and gain of the waveform), the amplifier outputs were modeled using the old response functions.
The inducted current waveform was continuously sampled at 2~MHz, although a specific DAQ system is under development, allowing for more flexibility in the sampling frequency. 

\item {\it Recirculation and purification}\\
The gas purity in terms of electronegative contamination is critical to ensure long electron drifts and reduce attachment, which impacts the final signal integral and, therefore, the energy resolution. For the R\&D, a system of hot and cold getters was used together with a recirculation piston pump. A purity allowing for an electron lifetime of $\sim$2~ms was achieved at 10~bar~\cite{Bouet:2024njk}, which is a factor of 5 worse than what is currently achieved in liquid xenon experiments~\cite{Plante:2022khm}. The system used was also limited in terms of gas flow (10 l/min) and maximum operating pressure (10~bar). For the proposed detector, the existing technology developed for noble liquid gas detectors could be used ({\it i.e.} cryogenic filters built from copper~\cite{Plante:2022khm} or the spark discharge techniques~\cite{Akimov:2017gxm}), granting a sufficiently pure gas to operate the proposed technology with a detector having the planned radius of 50~cm. The recirculation pump, a source of electronegative impurities, will also be replaced by a magnetically driven piston pump~\cite{LePort:2011hy}.
\end{itemize}

To install the detector in an underground laboratory, a footprint of approximately 50~m$^2$ is required to ensure safe operation of the detector and the recirculation/recuperation system. 

\section{Signal waveform construction}
\label{Sec:WF}
The sensitivity study relies on the detector simulation tools that were developed based on the best existing knowledge, exploiting the results of the detector response recorded during the entire R2D2 R\&D.

The physics simulation relies on GEANT4~\cite{GEANT4:2002zbu}: for all particles (signal and background) traversing the active volume, the energy deposits were saved. For each deposit, the position and the released true energy were recorded, forcing a step size of 100~$\mu$m.
\begin{figure}[t]
    \centering
    \includegraphics[width=\columnwidth]{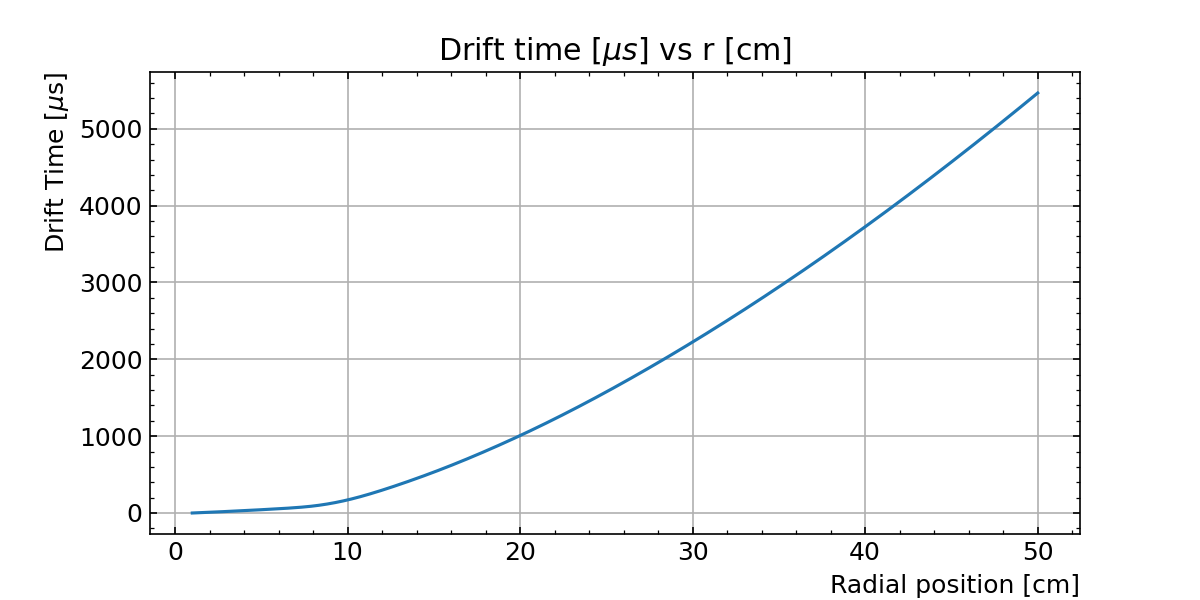}
    \caption{GARFIELD++ simulation of the drift time as a function of the radial position for a cathode of 50~cm at -20~kV and a central grounded anode with a radius of 1~cm. The CPC is filled with xenon gas at 40~bars.}
    \label{Fig:Dt}
\end{figure}
\begin{figure}[t]
    \centering
\includegraphics[width=\columnwidth]{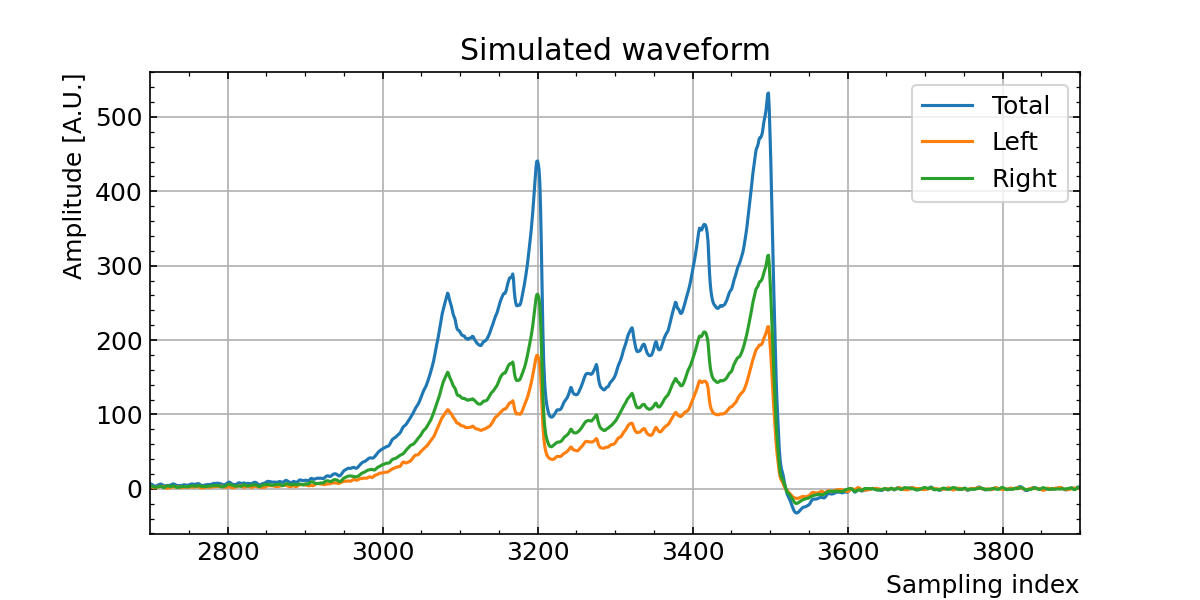}
    \caption{Example waveform of a $\beta\beta0\nu$ events starting at a radial position of 26.5~cm and a longitudinal position of Z=139~cm. The start (not shown) of the signal begins at bin 0. Each time sample corresponds to 0.5~$\mu$s given the sampling at 2~MHz.}
    \label{Fig:WFexample}
\end{figure}
The second step of the simulation  involves constructing the registered waveform based on the GEANT4  information. To do this, the procedure is divided into three steps:
\begin{itemize}
\item The energy deposited is converted into a number of ionisation electrons that are then  drifted to the anode. This conversion is achieved by dividing the deposited energy by the energy required to create an electron-ion pair ({\it i.e.} the so-called $W_i$), which is 22~eV in xenon~\cite{NEXT:2012vuv}. An integer number of electrons is assumed, although no additional smearing or quenching are accounted for. Previous studies on quenching~\cite{NEXT:2012vuv} suggests indeed that it can be neglected, whereas energy smearing based on experimental results will be applied at the analysis level (see Sec.~\ref{Sec:ERec}).
\item Using a GARFIELD++~\cite{Veenhof:1993hz} simulation of the detector geometry, the electron drift time to reach the anode is computed for each energy deposit and used to construct the waveform. 
A high voltage of -20~kV is applied to the cathode at a radius of 50~cm, while the central anode with a radius of 1~cm is grounded. The resulting drift time as a function of the starting electron radial position is shown in Fig.~\ref{Fig:Dt}. Validation of the GARFIELD++ simulated drift time against data was performed as part of the R\&D work, demonstrating reasonable agreement~\cite{Bouet:2022kav}.
\item Based on the Shockley-Ramo theorem (see Ref.~\cite{Bouet:2024njk} for details), two partial waveforms are reconstructed for each energy deposit: $Sl$ for the left side and $Sr$ for the right side of the anode. 
These two time series originate from the same initial partial waveform $Si$, which is truncated according to the position $R$ of the deposit. However, their amplitudes differ depending on the charge distribution, which is affected on the longitudinal position $Z$ of the deposit in the simulation. Assuming an anode resistance of several k$\Omega$, amplifier input impedances of the order of $\Omega$~\cite{Cuss:2002}, and taking $Z=0$ at the left edge of the detector, they conform to the following relationships:
\begin{equation}
\label{eq:nobg}
Sl = Si \times~(1-Z/Z_{det})
\end{equation}
\begin{equation}
Sr =  Si \times~(Z/Z_{det})
\end{equation}
where $Si$ represents the partial waveform without charge-sharing effect ({\it i.e.} $Si = Sl + Sr$), 
$Z$ is the GEANT4 hit position along the anode axis, and $Z_{det}$ is the total anode length (2.5~m). 
After adding all partial contributions from energy deposits, a convolution with an electronic response and filtering is applied. To mimic real noise, noisy snapshots based on data are independently added to $Sl$ and $Sr$. These time series were extracted from the measurements presented in Ref.~\cite{Bouet:2024njk}. To reflect real conditions, the true total signal $S$, used later in the analysis, is then constructed by summing $Sl$ and $Sr$, which contain uncorrelated noise sequences.
\end{itemize}

An example of the waveform of a $\beta\beta0\nu$ event is shown in Fig.~\ref{Fig:WFexample}.

\section{Waveform analysis}
\label{Sec:WA}
\begin{figure}[tp]
     \centering     
   	 \subfigure[\label{fig:Wdecomp1}]{\includegraphics[width=\columnwidth]{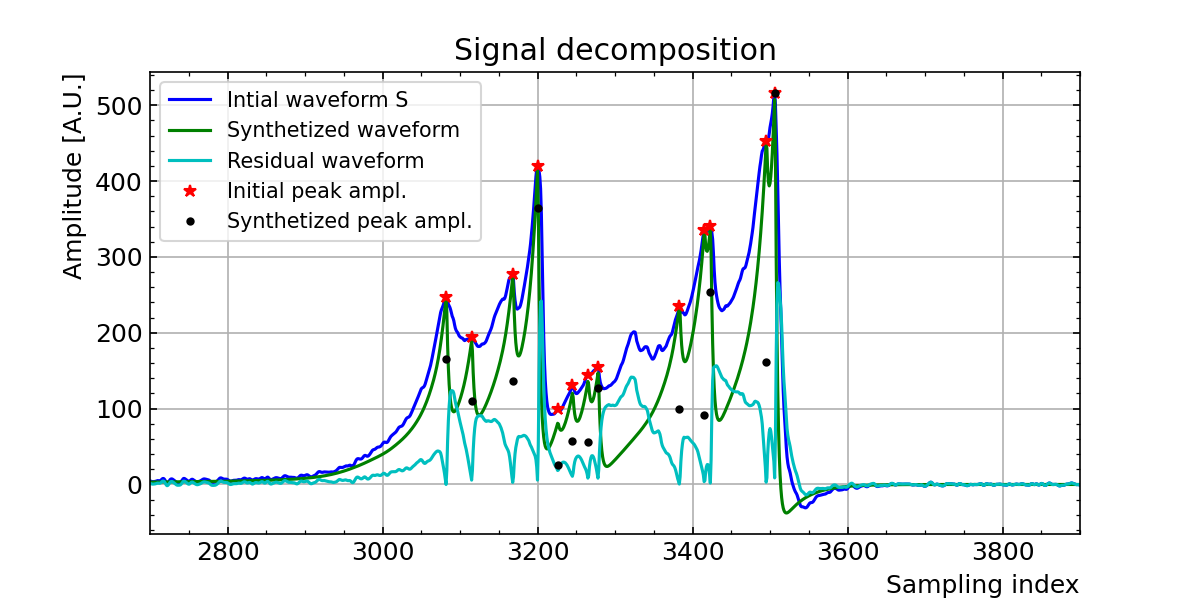}}
      \subfigure[\label{fig:Wdecomp2}]{\includegraphics[width=\columnwidth]{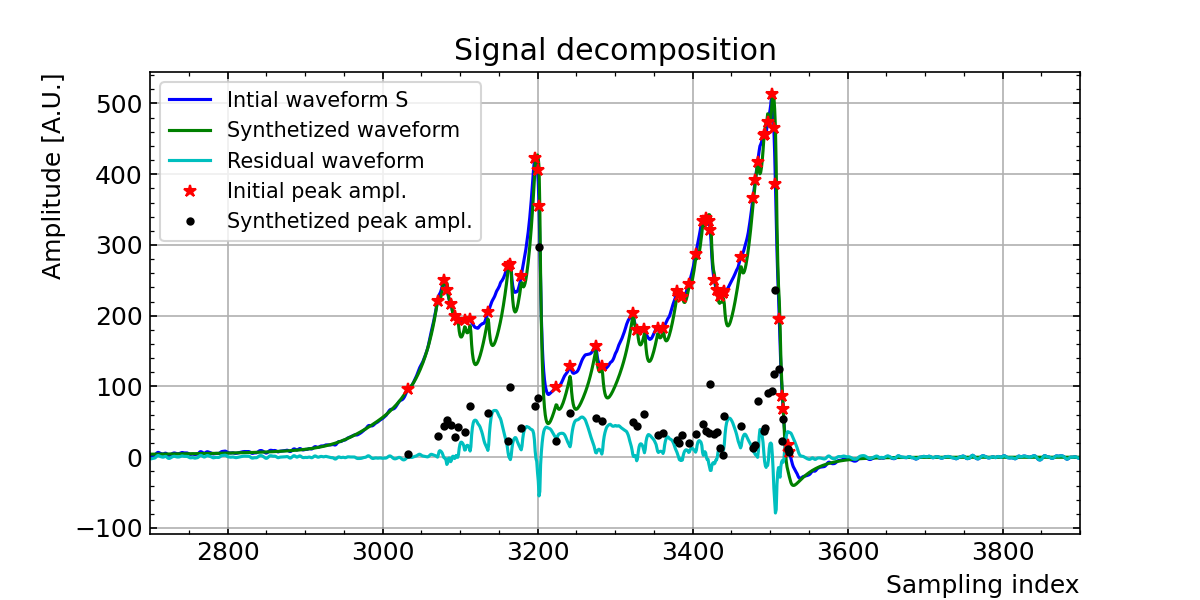}}
    \caption{Example of the decomposition of the total signal $S$ of the same $\beta\beta0\nu$ simulated event shown in Fig.~\ref{Fig:WFexample}. The result after the first iteration is shown in Fig.~\ref{fig:Wdecomp1}, whereas the result at the end of the process is presented in Fig.~\ref{fig:Wdecomp2}.The blue line indicates the 
original waveform, the green line shows the decomposition result, and the cyan curve displays 
the residue of the decomposition. The red stars mark the original amplitude of the different elementary 
contributions identified by the decomposition algorithm, whereas the black dots represent the amplitude of the contributions after the decomposition.}
    \label{Fig:Wdecomp}
\end{figure}
Once the anodic signals have been generated, it is necessary to develop a method for analyzing them. Following the procedure developed in previous works~\cite{Bouet:2024njk}, some global observables can be derived using the total waveform $S$, namely:
\begin{itemize}
\item The total reconstructed charge $Q_r$. An overall calibration is performed to align the peak of $\beta\beta0\nu$ events corresponding to the expected $Q_{\beta\beta}$ of 2.458~MeV.
\item The total signal duration $D_t$.
\end{itemize}

The analysis of the internal modulation of the signals can be performed to recover the radial and longitudinal location of the energy deposits, and thus reconstruct the 
particle track inside the gas. 
Due to the spiked shape of the induced current in the radial TPC 
(see Fig.~\ref{fig:Elost} in Sec.~\ref{Sec:ERec} and Ref.~\cite{Bouet:2024njk}), a local maximum in the signal is generated 
each time a primary reaches the anode (or, within the limits of the 
time resolution, a cluster of primaries). In this approach, the amplitude 
of each peak is directly related to the amount of charge reaching the anode. 
Regarding radial localization, the sampling index of the peak, subtracted from that of the signal’s start (common to all peaks in the same event), determines the drift time associated with the peak (calculated by multiplying by the sampling period). The correspondence between this time lapse and the radial position of the energy deposition for the primary is provided in Fig.~\ref{Fig:Dt}, enabling the extraction of a radial position $R$ for the peak. Consequently, referencing an external time stamp is no longer necessary.
The longitudinal position $Z$ of a peak is determined by the principle of charge sharing, using the ratio of the left ($S_l$) and right ($S_r$) amplitudes of the peak, as described in Sec.~\ref{Sec:WF}.

Building upon previous calculations, a complete decomposition of the total signal $S$ into elementary contributions of induced current has been developed. This process is achieved (see Fig.~\ref{Fig:Wdecomp}) using a simple iterative method for signal decomposition and synthesis. 
The decomposition involves identifying peaks in the original waveform and scaling the elementary current template (truncated if necessary) in amplitude to match each peak. The synthesized waveform is then formed by summing these scaled contributions. 
Any remaining peaks not previously identified are searched for in the residual waveform 
resulting from the subtraction of the synthesized waveform from the original one. The process continues iteratively until the ratio of the synthesized charge to the original charge, $Q_{synthesized}/Q_r$, exceeds 0.99. 

\begin{figure}[t]
    \centering
    \includegraphics[width=\columnwidth]{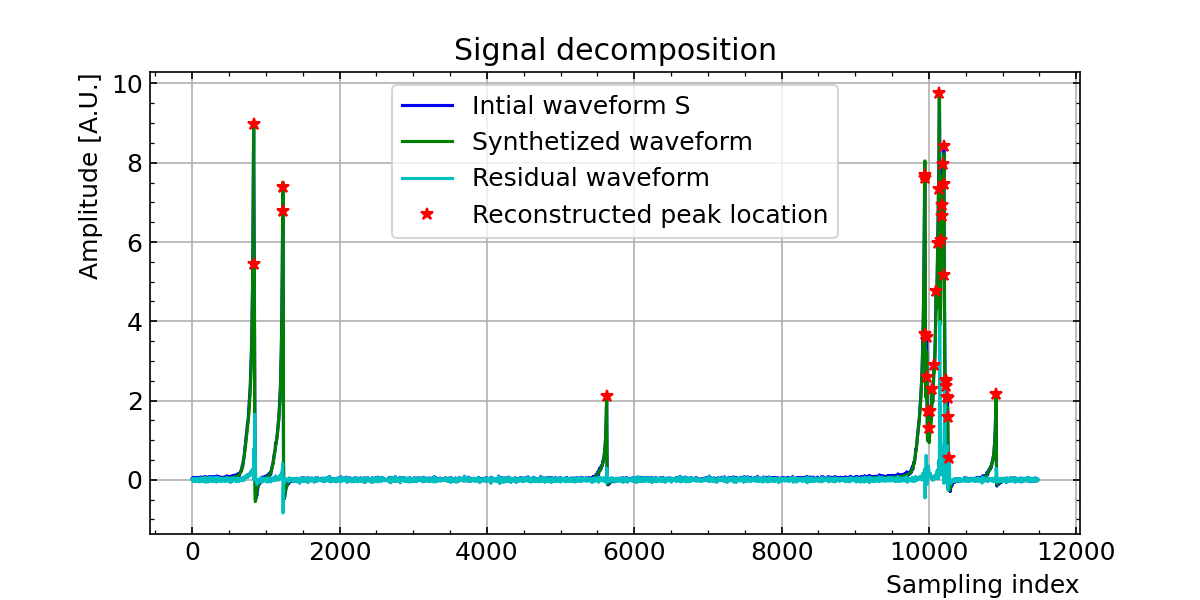}
    \caption{Example of a signal decomposition (in full time range) showing multiple distant interactions, originating from a single $^{208}$Tl event produced by the simulation of the $^{232}$Th decay chain.}
    \label{Fig:ThWdecomp}
\end{figure}

\begin{figure}[tp]
    \centering
    \includegraphics[width=\columnwidth]{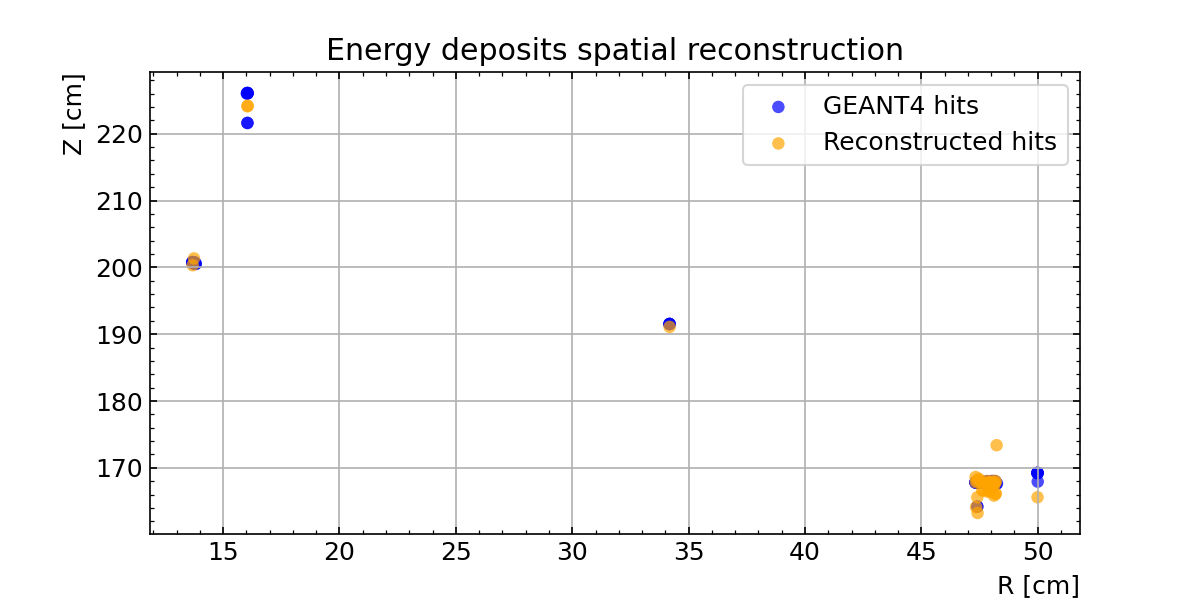}
    \caption{Reconstruction of the elementary interactions (orange points) for the same $^{208}$Tl 
event shown in Fig.~\ref{Fig:ThWdecomp}, in the natural $(R, Z)$ plane as induced by the radial TPC. 
The blue points correspond to the original GEANT4 hit locations.}
    \label{Fig:Hitloc}
\end{figure}

The main advantage of this method is that it allows for a more accurate and detailed 
map of the interaction footprint in the $(R,Z)$ plane (see Fig.~\ref{Fig:ThWdecomp} and~\ref{Fig:Hitloc}). 
Furthermore, the decomposition allows for extracting the true amplitude of each peak, making it possible to estimate the actual number of primaries associated with each. Finally, through synthesis, the unobserved portion of the signal, which depends on the initial radial position of each primary, becomes quantifiable. This enables the reconstruction of the true energy of the signal, determined by integrating both the observed and unobserved (lost) signal components (see Sec.~\ref{Sec:ERec}).

Ultimately, this approach allows the extraction of the following ancillary observables, which have been used for further analyses:
\begin{itemize}
\item The number of peaks ($N_p$) in the waveform . As explained before, for each peak, 
the radial position is reconstructed based on its drift time, and its $Z$ position is 
determined from the ratio between the left and right waveforms.
\item The event radial elongation $\Delta R$ is computed as the maximal distance between the radial positions of the individual peaks.
\item The event longitudinal elongation $\Delta Z$ along the $Z$ axis corresponds to the 
maximal distance between the $Z$ positions of the individual peaks.
\item The largest radial position of the reconstructed peaks, denoted as $MaxR$.
\end{itemize}

\begin{figure}[tp]
    \centering
    \includegraphics[width=\columnwidth]{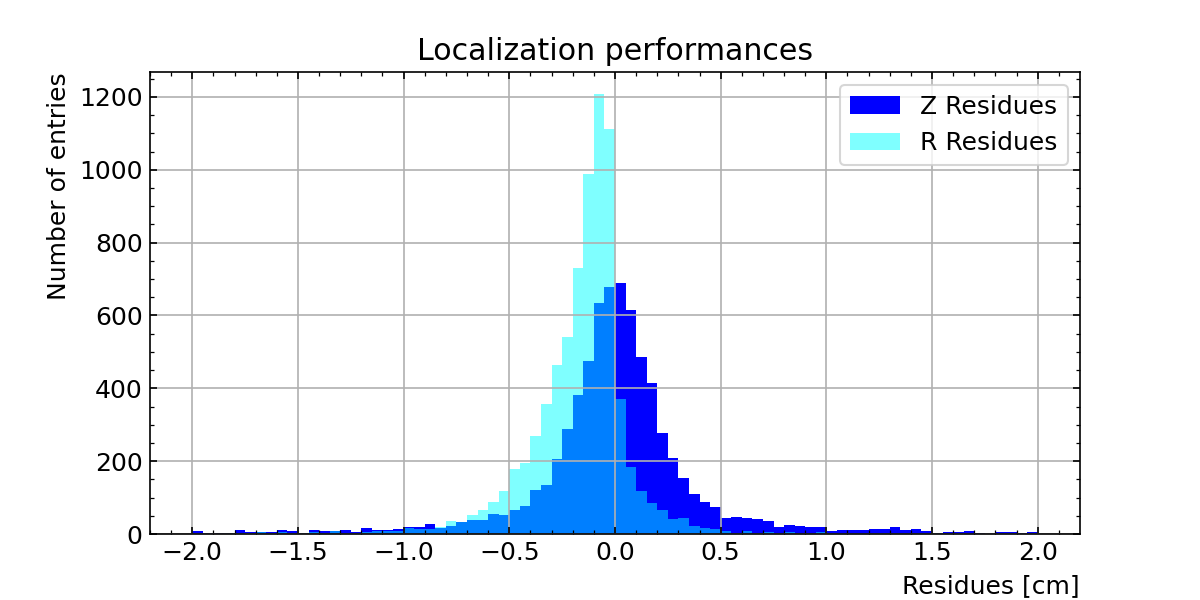}
    \caption{Residual distributions in $R$ and $Z$ for the simulated signal interactions, including satellite deposits  resulting from the multiple interactions in the gas.}
    \label{Fig:Residu}
\end{figure}

Associated with specific cuts discussed in Sec.~\ref{Sec:Cuts}, these estimators play a 
key role in distinguishing the topology of the $\beta\beta0\nu$ signals from the background.
The quality of the localization of the interactions was estimated by comparing the barycenters in $R$ and $Z$ (computed using the charge of the deposits as weight) for the GEANT4 events with those reconstructed by our processing. The distributions of the residuals are presented in Fig.~\ref{Fig:Residu}).

Although the present simulation did not account for either the lifetime or the diffusion of the primaries, the most evident effects induced by these contributions on the simulated waveforms were examined.
The reduction in the number of primaries during their migration was modelled according to the relation:
\begin{equation}
\label{eq:att}
N(t) = N(0) e^{-\frac{t}{\tau}},
\end{equation}
where  $\tau$  is the lifetime of the electrons,  $N(0)$  represents the number of primaries created at time  $t$ = 0  prior to their drift, and  $N(t)$  is the number of primaries remaining after a drift time  $t$.
For each hit, the associated charge loss is accounted for by multiplying the truncated form of this function with the corresponding elementary current.

As a byproduct of the GARFIELD++ results presented in Fig.~\ref{Fig:Dt}, the spatial dispersion of the primaries (represented as the standard deviation of the distribution under the Gaussian hypothesis) was also extracted as a function of the distance travelled. This spatial dispersion was then converted into a temporal dispersion, $\sigma_d$. The latter, considered as an overall diffusion effect, was found to increase almost linearly with distance, starting from $\sigma_d = 0~\mu$s for primaries generated near the anode, up to $\sigma_d \approx 4.2~\mu$s for those generated near the cathode. The complete waveform, incorporating diffusion effects, was obtained by convolving each elementary current function with a Gaussian whose standard deviation corresponds to the starting point of the hit.

\begin{figure}[t]
    \centering
    \includegraphics[width=\columnwidth]{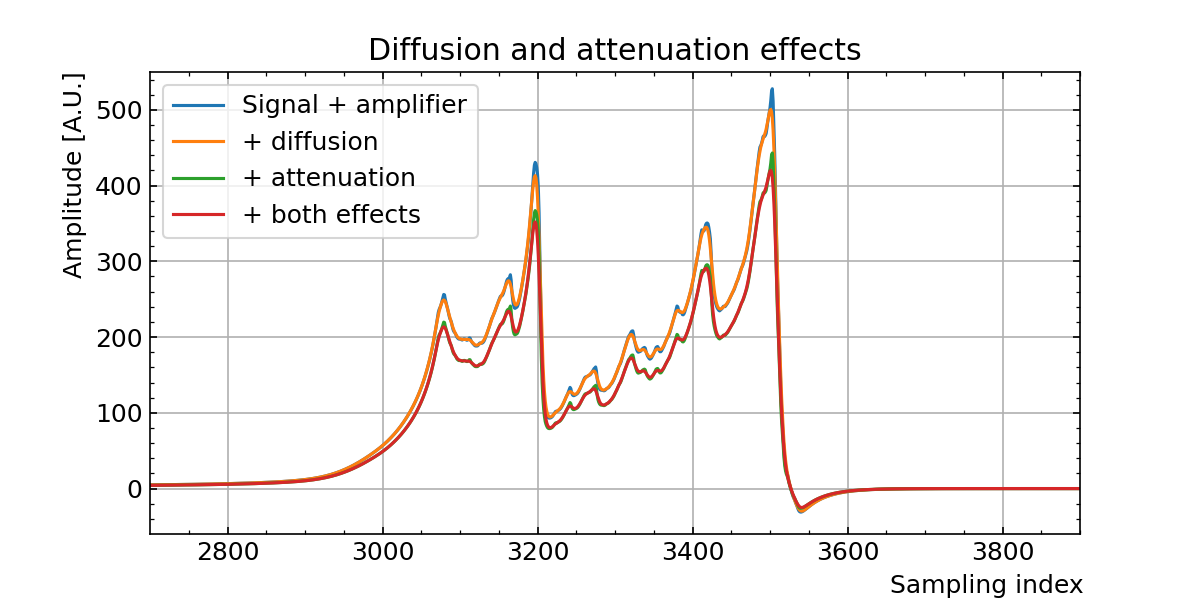}
  \caption{Signal deformation under the combined effects of attenuation and diffusion, assuming $\tau = 10~\mathrm{ms}$ and $\sigma_d = 2.7~\mu\mathrm{s}$ (at $R = 26.5~\mathrm{cm}$).}
    \label{fig:Effect}
\end{figure}

For illustration, the deformation of the waveform shown in Fig.~\ref{Fig:WFexample} is presented in Fig.~\ref{fig:Effect}, with parameters $\tau = 10~\mathrm{ms}$ and $\sigma_d = 2.7~\mu\mathrm{s}$ (at $R = 26.5~\mathrm{cm}$). Further studies as a function of the interaction distance suggest that attenuation is the most significant detrimental effect. However, the signal can still be effectively decomposed as long as $\tau$ remains greater than the maximum drift time. These preliminary observations motivated the initiative not to include these two effects at this stage of the study.

\section{Energy reconstruction}
\label{Sec:ERec}
\begin{figure}[tp]
    \centering
    \includegraphics[width=\columnwidth]{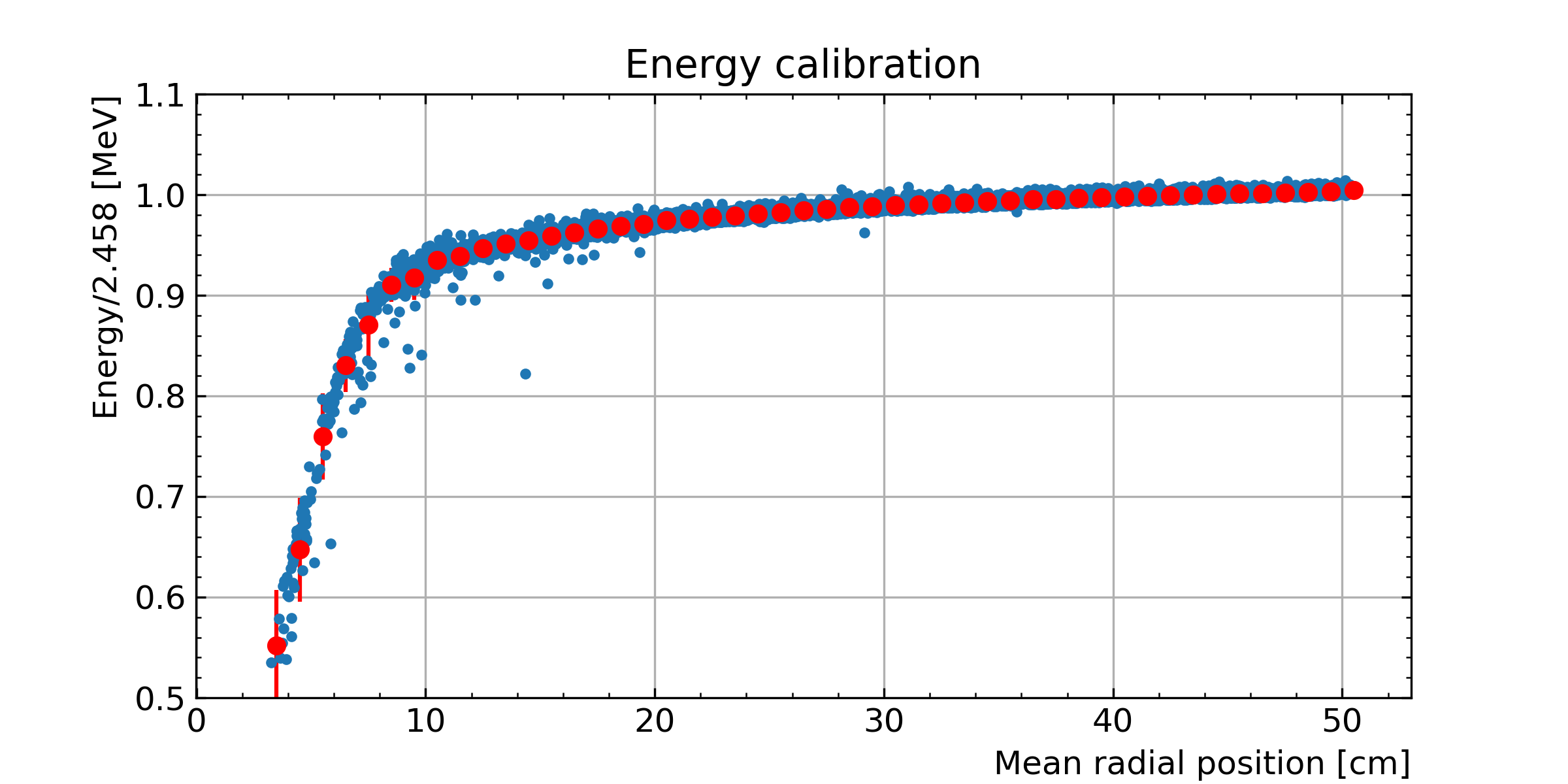}
    \caption{Fraction of energy with respect to the $Q_{\beta\beta}$ as function of the event mean radial distance (weighted by the charge for the different energy deposits). The red points represent the mean values with steps of 1~cm.}
    \label{fig:Calib}
\end{figure}
Energy reconstruction is critical as it directly impacts event selection and the definition of the so-called ROI ({\it i.e.} Region Of Interest).

\begin{figure}[tp]
    \centering
\includegraphics[width=\columnwidth]{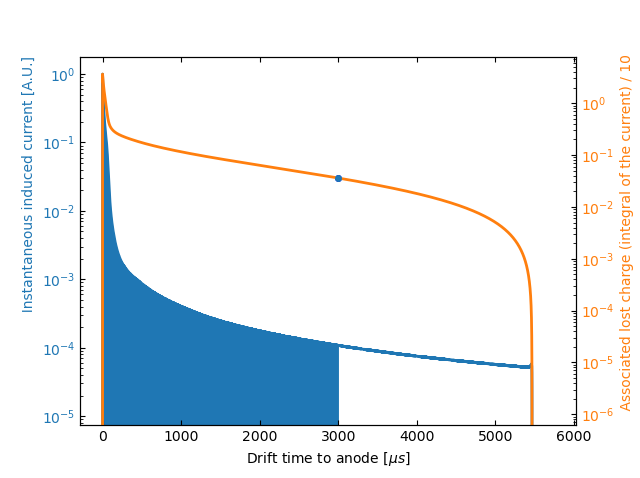}
    \caption{Instantaneous current (blue) induced by an elementary charge and corresponding lost charge (orange) as a function of the drift time. For illustration, the marker indicates the energy loss associated with a primary created at $R$ = 35.38~cm, and the corresponding waveform is shown by the blue area.
    }
    \label{fig:Elost}
\end{figure}

\begin{figure}[t]
    \centering
    \includegraphics[width=\columnwidth]{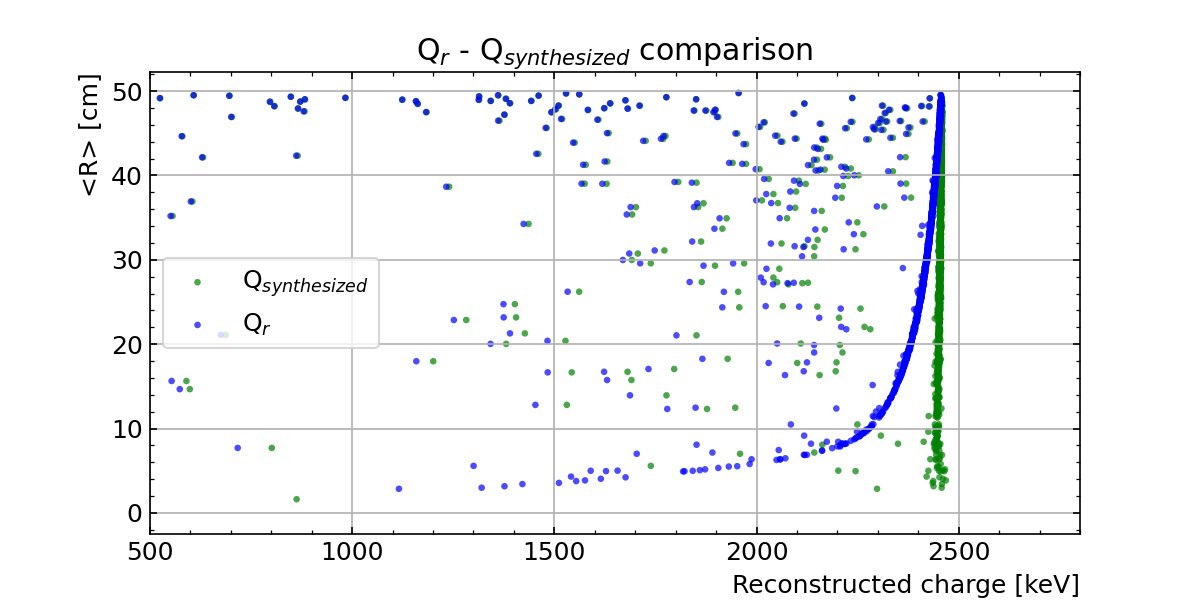}
   \caption{Reconstructed charge $Q_{r}$ (without the lost charge)  and $Q_{synthesized}$ (including the lost charge) as a function of the signal mean radial position $R$ when no calibration procedure is applied (simulated $\beta\beta0\nu$ events).}
    \label{Fig:Qcomparison}
\end{figure}

As mentioned in Sec.~\ref{Sec:WA}, the waveform integral is normalised to match the signal $Q_{\beta\beta}$; however, the reconstructed charge depends on the radial position of the energy deposit, since only drifted electrons contribute to the observed signal, while ions are below the noise level due to their low mobility (see Ref.~\cite{Bouet:2024njk} for more details). For this reason, the true energy must be estimated relative to a common reference, by correcting the deficit associated with charge produced at radial position $R$ compared to that at the cathode. Two correction strategies were developed:
\begin{itemize}
\item The reconstructed energy as a function of the radial position was simulated for $\beta\beta0\nu$ events. This robust approach is based on a correction of the signal integral as a function of the mean radial distance R of the trace. As illustrated in Fig.~\ref{fig:Calib}, where the red dots represent the mean values in 1~cm intervals, a linear interpolation between these points is used to derive a corrective function, which calibrates the apparent energy as a function of the radial position $R$ of the events.
\item The synthesis method also allows calculation of the reconstructed energy deficit. For each cluster of an event, identified by the decomposition algorithm at a given radial position $R$ ({\it i.e.} at a given drift time), the energy deficit is determined by multiplying the value of the lost charge function at that position by the number of primaries estimated at the same point. Summing over the different clusters enables the correction of the energy deficit due to the varying initial radial positions of the primaries in the event (see Fig.~\ref{fig:Elost}). To demonstrate the effectiveness of the method, Fig.~\ref{Fig:Qcomparison} presents a comparison between the total reconstructed charge, $Q_r$, and the reconstructed charge after applying the decomposition algorithm, $Q_{synthesized}$, without calibration. The comparison is shown as a function of the mean radial position of the signal.
\end{itemize}

\begin{figure}[t]
    \centering
    \includegraphics[width=\columnwidth]{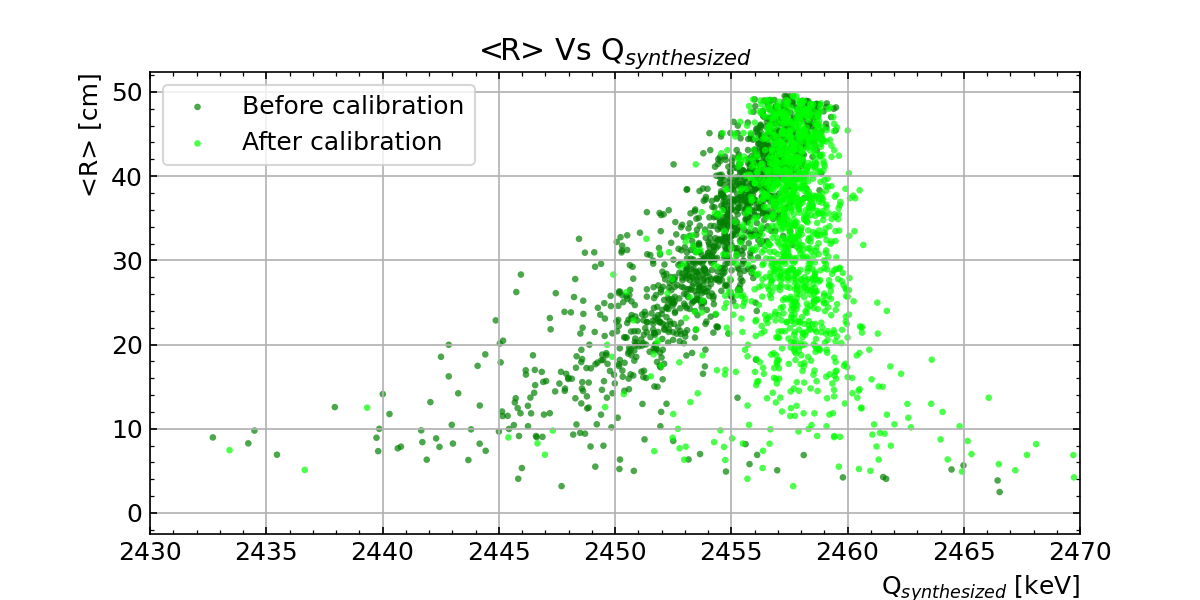}
   \caption{Reconstructed charge $Q_{synthesized}$ from the synthesis method as a function of the mean radial position of the signal $R$ before and after the calibration procedure explained in the text (simulated $\beta\beta0\nu$ events).}
    \label{fig:QsR}
\end{figure}

Although the synthesis method does not require additional calibration in principle, a weak radial dependence was still observed in the identification diagram ($Q_{synthesized}$, $R$) (see Fig.~\ref{fig:QsR}). This phenomenon can be attributed to errors in defining the signal origin ({\it i.e.} when the signal crosses the threshold), which are influenced by noise and have a more significant impact when the signal originates near the anode, where the signal slope is steeper. Therefore, a second corrective step was implemented, utilizing the previously described method applied to the signal integral, to further enhance the calibration.

The study highlights the critical importance of experimental energy calibration of the detector due to the dependence of the signal on the radial position $R$. This calibration must also be implemented in real operation, and two potential solutions are currently under investigation.
The first option involves using a dedicated point-like alpha source deployed at specific radial positions, though this could potentially disturb the drift electric field. Another attractive solution is to exploit the diffuse radon dissolved in the gas. 

\begin{figure}[t]
    \centering
    \includegraphics[width=\columnwidth]{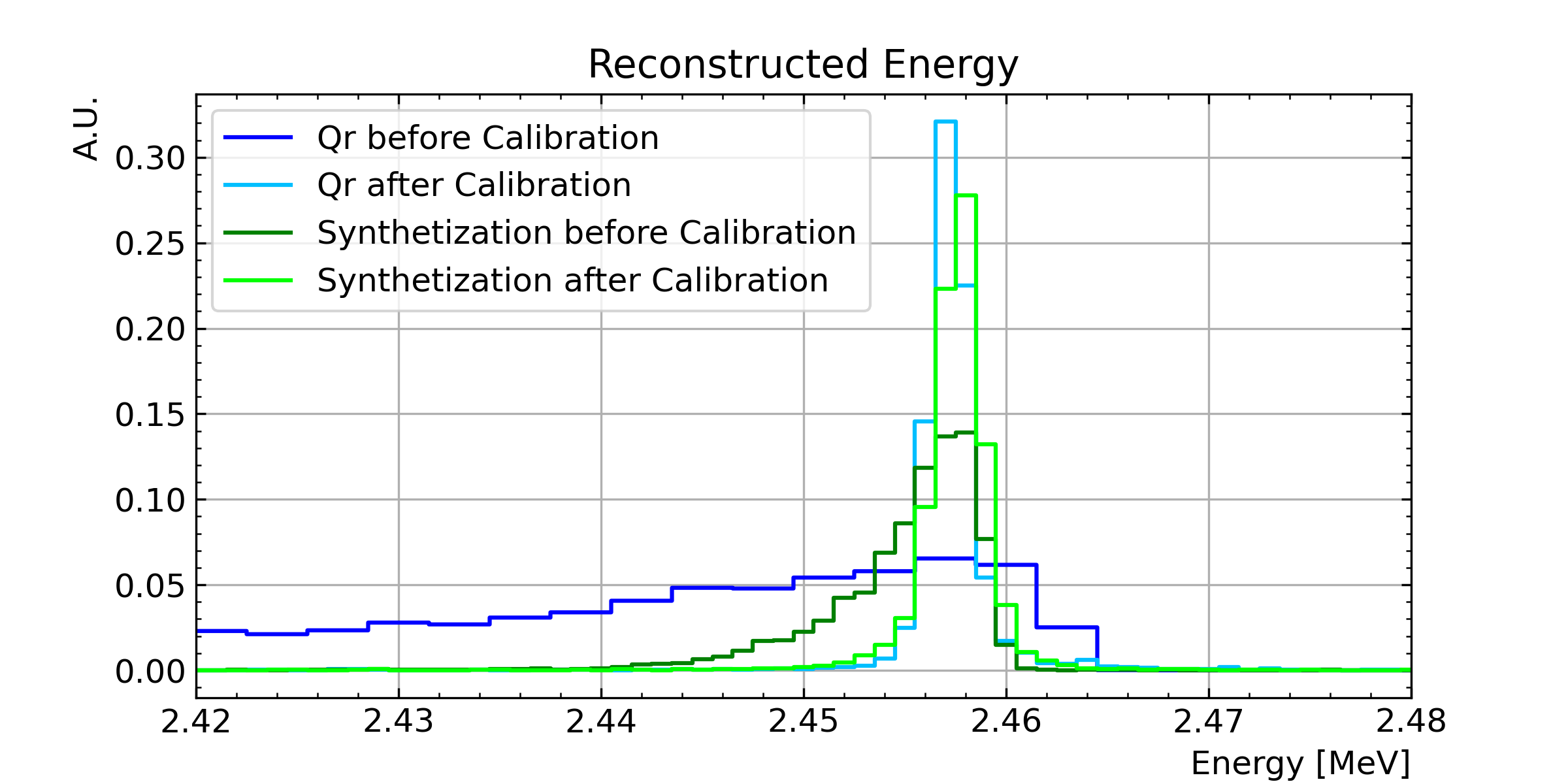}
  \caption{Energy reconstructed by integrating the simulated signal waveform of $\beta\beta0\nu$ events before (blue line) and after (cyan line) the calibration procedure. The results after the waveform synthesis are also shown, both before (green line) and after (light green line) the calibration procedure.}
    \label{fig:Ene}
\end{figure}

The energies reconstructed before and after the calibration procedures previously discussed are shown in Fig.~\ref{fig:Ene}. A Gaussian fit of the distribution obtained after energy correction shows a resolution of 0.1\% FWHM for both methods (intrinsic resolution of the method since stochastic fluctuation on the number of created charges is not included at this stage). Although both approaches yield similar results, the synthesis routine is deemed more satisfactory from a methodological perspective. Consequently, this method was used for energy estimation in this article. 

To perform a realistic sensitivity study, the energy resolution must account for additional uncertainties. At this stage, effects due to fluctuations in the number of electrons or charge loss from attachment to gas impurities during drift have not been included. A separate statistical error of 0.3\% FWHM is expected, assuming a Fano factor of approximately 0.15~\cite{Nygren:2011zz}. Additionally, for real data, the influence of front-end electronics (amplifier response function) or digital processing (time shift of the sampling comb between the two ADCs) can also affect the energy resolution. It should be noted that these effects impact only the reconstructed energy and not 
the efficiency of the peak-finding algorithm or the selection variables derived from it. 
This was verified through a dedicated simulation of waveforms, which included a stochastic 
fluctuation in the number of electrons produced for each energy deposit.

Previous experimental work on proportional parallel projection chambers achieved a resolution of 0.7\% FWHM at 660~keV~\cite{BOLOTNIKOV1997360}. When rescaled as $1/\sqrt{E}$ to the xenon $Q_{\beta\beta}$, this corresponds to a resolution of 0.4\% FWHM, close to the statistical limit. Experimental results from the R2D2 CPC showed an energy resolution of 1.2\% FWHM at 5.3~MeV~\cite{Bouet:2024njk}. Optimised analysis based on the signal processing discussed in Sec.~\ref{Sec:WA} enabled a resolution of 0.7\% FWHM at 5.3~MeV, which corresponds to 1\% FWHM at the xenon $Q_{\beta\beta}$, the original goal of the R2D2 R\&D. The factor-of-two difference compared to Ref.~\cite{BOLOTNIKOV1997360} can be attributed to the experimental results being dominated by electronic noise and gas quality. Future experimental results are expected to improve significantly with better noise conditions, upgraded electronics and better-purified gas.

\begin{figure*}[tp]
    \centering     
   	 \subfigure[\label{fig:cut1}]{\includegraphics[width=\columnwidth]{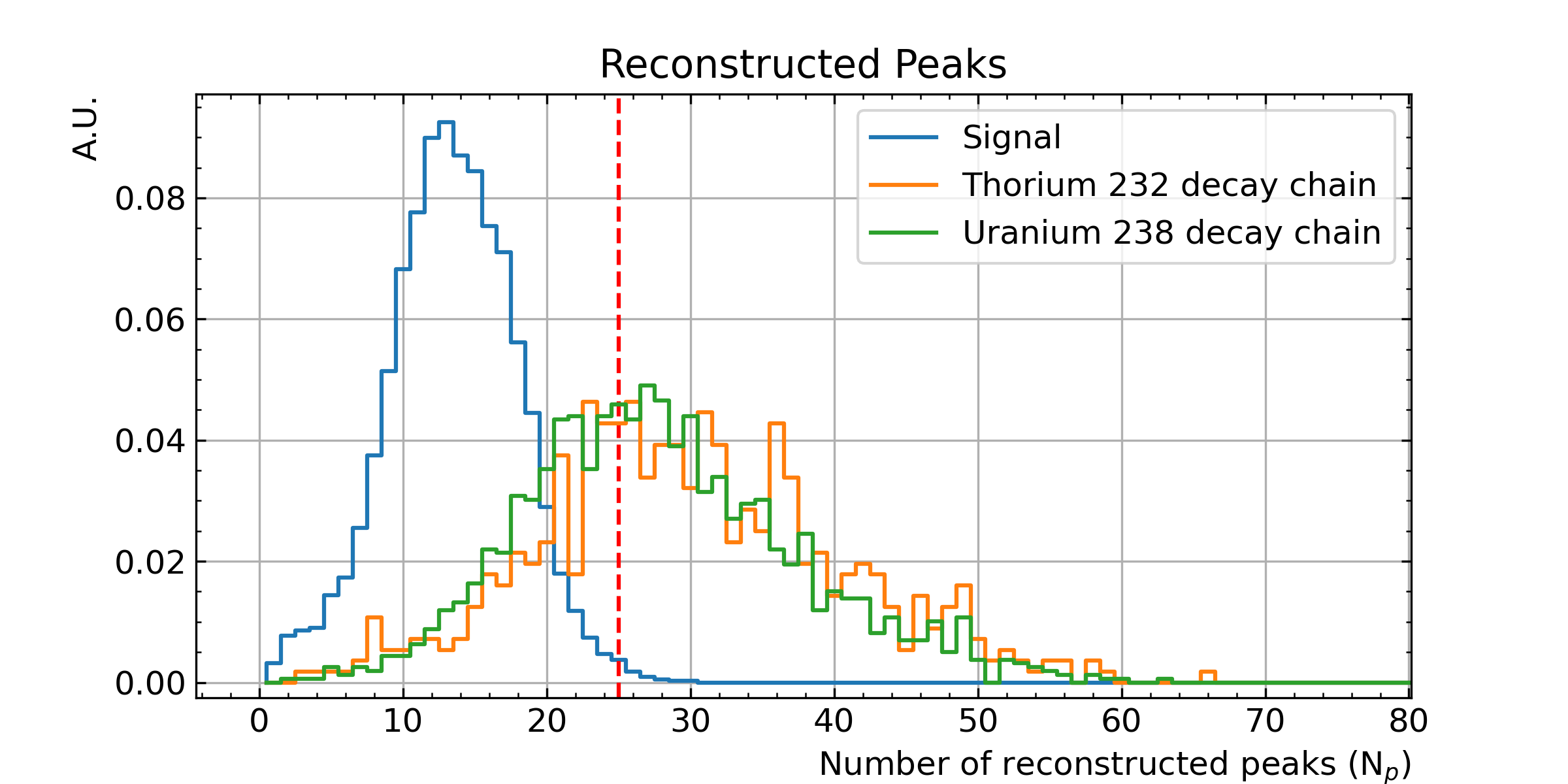}}
    	 \subfigure[\label{fig:cut2}]{\includegraphics[width=\columnwidth]{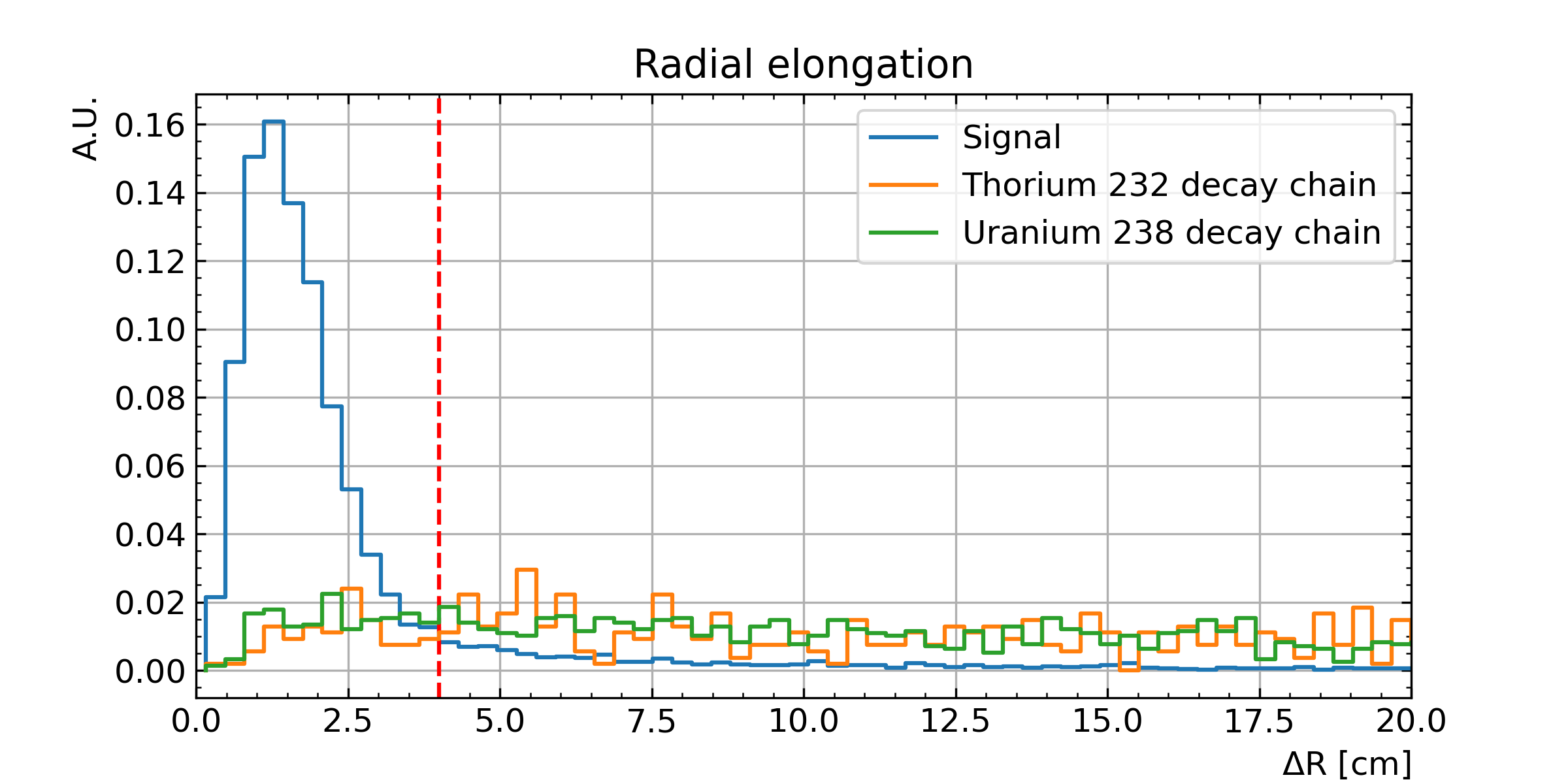}}
      	 \subfigure[\label{fig:cut3}]{\includegraphics[width=\columnwidth]{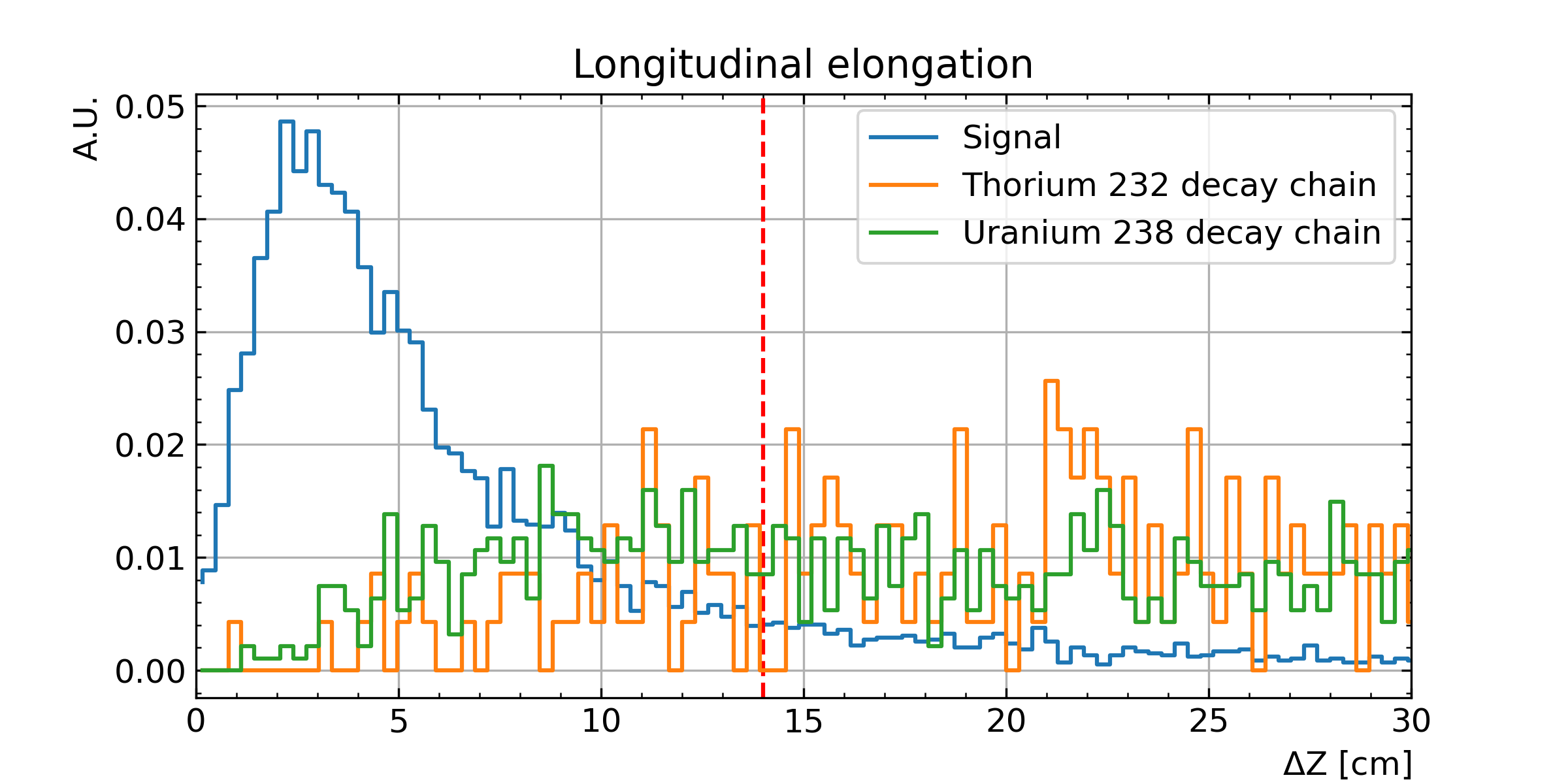}}
        	 \subfigure[\label{fig:cut4}]{\includegraphics[width=\columnwidth]{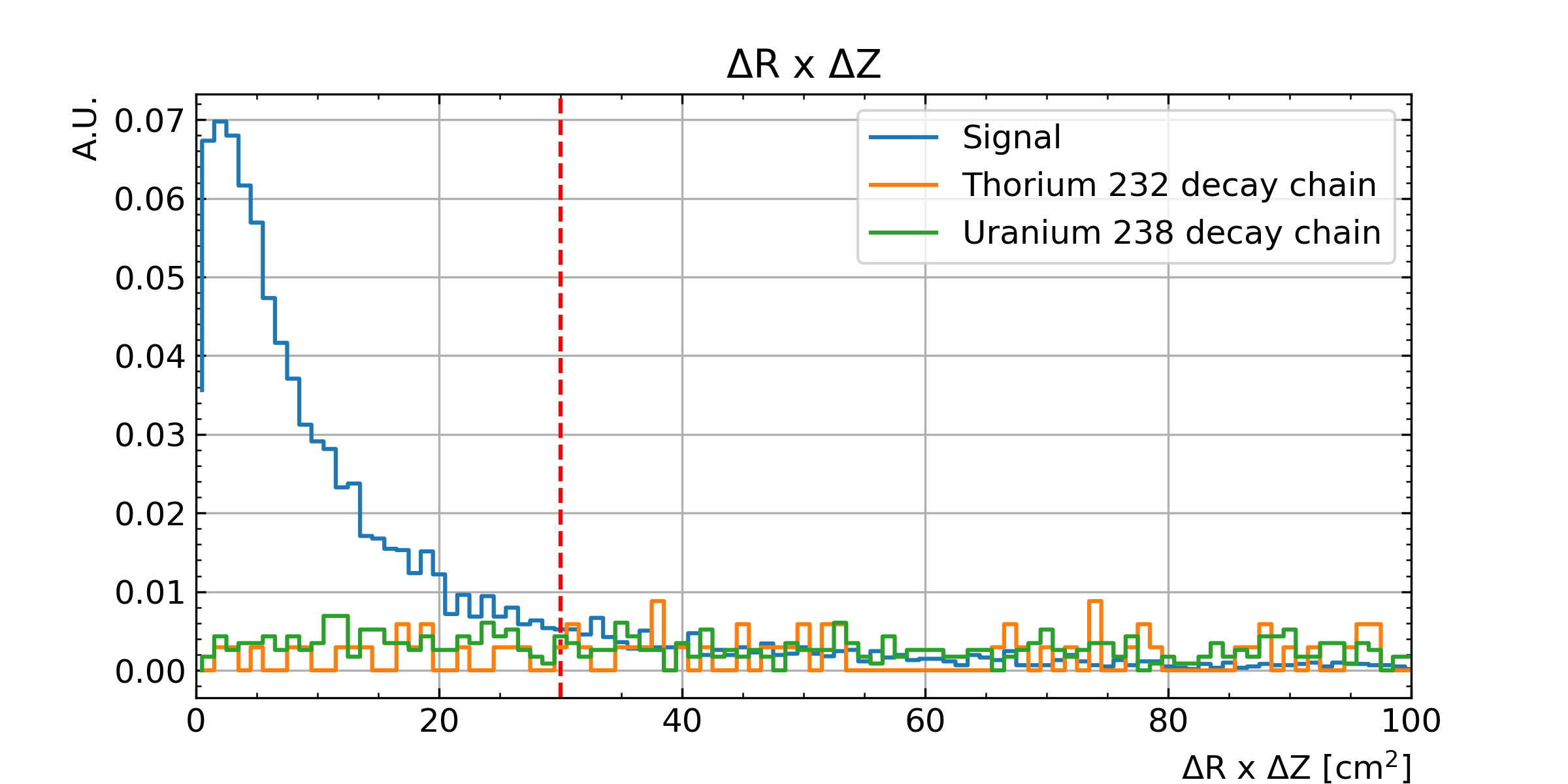}}
          	 \subfigure[\label{fig:cut5}]{\includegraphics[width=\columnwidth]{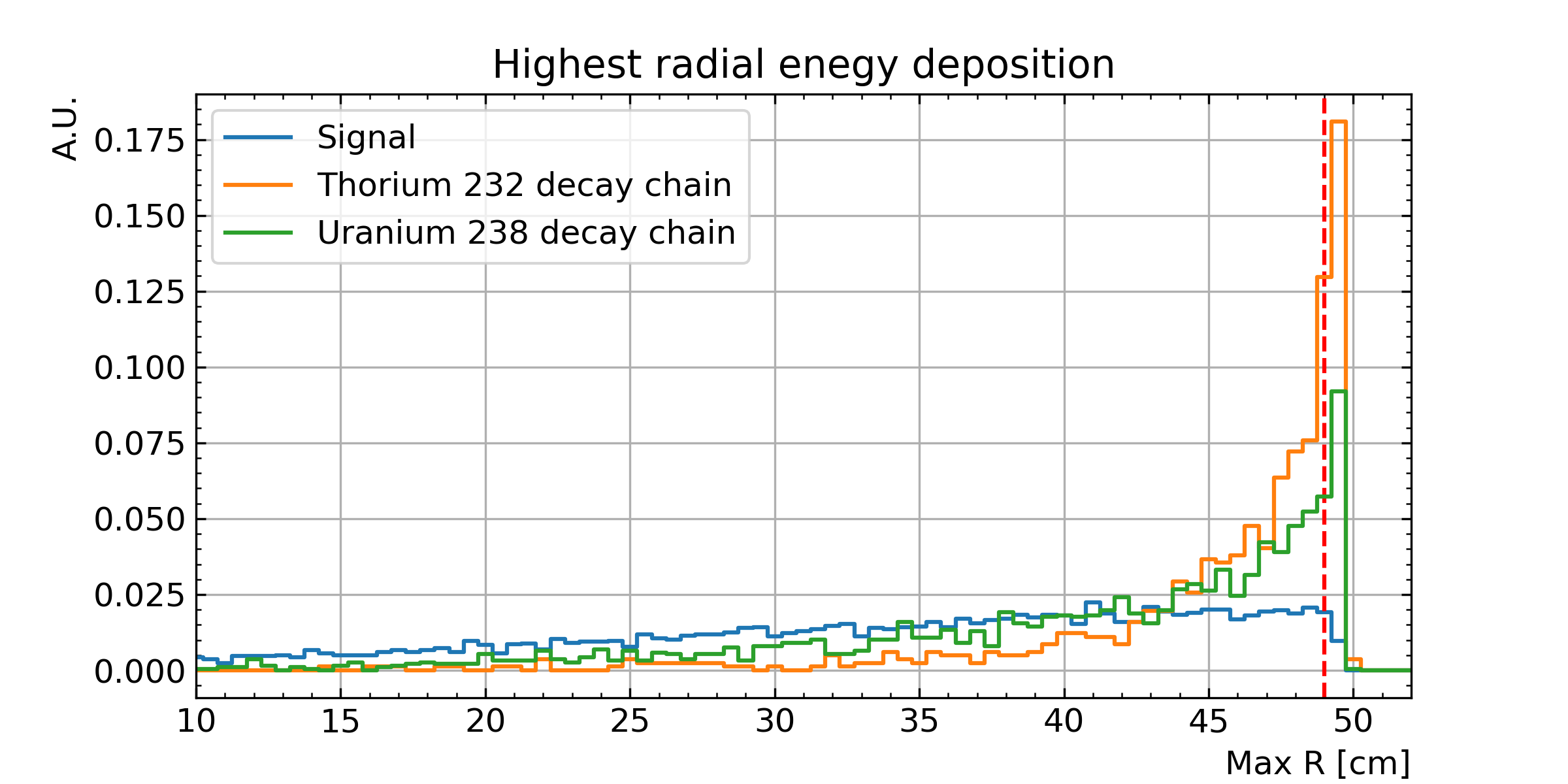}}
    \caption{Selection cuts, indicated by the vertical red dashed line, on several observables, namely: number of reconstructed peaks~\ref{fig:cut1}, $\Delta R$~\ref{fig:cut2}, $\Delta Z$~\ref{fig:cut3}, $\Delta R \times \Delta Z$~\ref{fig:cut4}, and larger radial energy deposit position~\ref{fig:cut5}. The distribution of such observables (after ROI selection) for $\beta\beta0\nu$ (blue), Thorium 232 chain (orange) and Uranium 238 chain (green) are shown.}
    \label{fig:Cuts}
\end{figure*}

To account for the aforementioned uncertainties, 
 an energy resolution of 1\% FWHM at the xenon $Q_{\beta\beta}$ is conservatively chosen for the following sensitivity study. Consequently, the reconstructed energy of each event is spread using a Gaussian function, where the width corresponds to this assumed final energy resolution.

\section{Selection cuts}
\label{Sec:Cuts}

The first selection is based on the event’s reconstructed energy, considering only those events that fall within the defined ROI. This ROI selection is closely related to the achieved energy resolution, and previous studies have shown that an ROI with a half-width equal to the energy resolution, expressed as FWHM, is the optimal choice~\cite{Meregaglia:2017nhx}. Given the assumed energy resolution of 1\% (see Sec.~\ref{Sec:ERec}) at the xenon $Q_{\beta\beta}$, this translates to an ROI in the range of [2.433 -- 2.483]~MeV.

Additional selection cuts are based on the observables built from the reconstruction. They are tuned to reduce the background in the ROI as much as possible while keeping the signal acceptance at a reasonable level of at least $\sim 60\%$. The distributions of these observables, associated with the typical cuts used for the $\beta\beta 0\nu$ signal and the main background components, are shown in detail in Fig.~\ref{fig:Cuts}. The selections on the different variables are detailed below with an explanation of the logic behind each choice.
It is important to note that, at this level, the geometric reconstruction of events depends only weakly on the energy resolution, despite the presence of smearing.

A first source of background originates from $\alpha$’s or $\beta$’s arriving from the vessel or from outside. To mitigate this background, a selection requiring $MaxR$ to be smaller than 49~cm is applied, removing events with energy deposits in the outermost centimeter of the active volume.

A second source of background arises from $\gamma$’s interacting within the active volume and depositing energy within the accepted ROI. However, these events typically undergo multiple Compton interactions, resulting in a topology that is less compact compared to the expected signal electron tracks. Multiple spatial cuts demonstrated effective background rejection, namely $\Delta R < 4$~cm, $\Delta Z < 14$~cm, and $\Delta R \times \Delta Z < 30$~cm$^2$. Additionally, these events exhibit a large number of reconstructed peaks, prompting the application of a cut on $N_p < 25$.

\section{Signal simulation}
To simulate the signal, {\it i.e.} the two electrons emitted in the $\beta\beta0\nu$ decay, pre-computed spectra were used~\cite{Kotila:2012zza}, as this decay is not included in GEANT4. The angular correlation between the two electrons was not considered, but its impact is expected to be a second-order effect.

The events are simulated isotropically throughout the xenon volume, but only events generated in the active volume ({\it i.e.}, about 75\% of the total events) are used to estimate the detection efficiency. The breakdown of selection efficiencies after the various selection steps can be found in Tab.~\ref{Tab:signal}, resulting in a final signal selection efficiency of 61\%.

\begin{table}[ht]
\begin{center}
\begin{tabular}{|c|c|}
\hline
Events in ROI   & + Topology cuts \\

\hline
82.2\% & 60.8\% \\
\hline
\end{tabular}
\caption{Fraction of signal in the ROI without and with topological selection cuts.}
\label{Tab:signal}
\end{center}
\end{table}%

\section{Background simulation}
\label{Sec:BG}
\subsection{$\beta\beta2\nu$}
\label{Sec:BGbb}
The high-energy tail of events from the $\beta\beta2\nu$ process could contribute to background if it falls within the ROI, as these events would be indistinguishable from the sought-after signal. Assuming a 582~kg mass of pure $^{136}$Xe (see Sec.~\ref{Sec:Det}) and a mean lifetime $T^{2\nu}_{1/2}$ of $2.165 \times 10^{21}$ years~\cite{ParticleDataGroup:2022pth}, the total expected number of decays per year is approximately 825000.  Considering an ROI of [2.433 -- 2.483]~MeV, and accounting for 1\% FWHM energy smearing at the xenon $Q_{\beta\beta}$ of 2.458~MeV, only a fraction of $9 \times 10^{-10}$ events would fall in the ROI. This corresponds to less than 0.001 events per year, allowing this background source to be completely neglected.

\subsection{Composite material radioactivity}
\label{Sec:BGCompo}
A specific composite material selection process is currently underway in collaboration with the IRT Jules Verne company. Various carbon fibres are being evaluated based on their radioactivity and mechanical strength to withstand pressure, alongside different epoxy glues to bind them. As mentioned in Sec.~\ref{Sec:Det}, a structure with a 1.5~cm thick carbon fibre-based composite material has been assumed. The decay chains of $^{238}$U and $^{232}$Th, were simulated uniformly within the carbon fibre layer.
For the sensitivity studies, it was assumed that material radioactivity at the level of 10~$\mu$Bq/kg can be achieved. While this remains a critical point yet to be fully demonstrated, promising results have been obtained through the partnership with IRT Jules Verne. In any case, the presented background event rates per year can be scaled linearly with respect to different material radioactivity levels.

In the GEANT4 simulation, a conservative time window of 10~ms was assumed to separate events in the decay chain, considering a maximal drift time of approximately 5~ms. Given the half-lives of 164~$\mu$s for $^{214}$Po (from the $^{238}$U chain) and 300~ns for $^{212}$Po (from the $^{232}$Th chain), Bi-Po decays are treated as a single event in the detector simulation.
For each decay chain, a sample of 10 million events was simulated. Only events depositing more than 2.2~MeV in the xenon active volume at the GEANT4 level were further processed. Based on the assumed activity of 10~$\mu$Bq/kg and the composite mass of approximately 330~kg, the generated statistics correspond to about 10~years of data for $^{232}$Th, and 7~years for $^{238}$U.

The total number of expected events per year in the ROI, before any selection cuts, can be computed, as well as the irreducible background events surviving the selection cuts described in Sec.~\ref{Sec:Cuts}. A summary is presented in Tab.~\ref{Tab:radioactivity}, which shows that the total number of background events from the composite vessel’s radioactivity is approximately 1.2 per year. All the selected events in the $^{232}$Th decay chain originate from $^{208}$Tl decays, while the events from the $^{238}$U decay chain are from $^{214}$Bi decays. In both cases, a $\gamma$ deposits all the energy in a confined region of the detector, with no multi-Compton topology to enable event rejection. 

\begin{table}[ht]
\begin{center}
\begin{tabular}{|c|c|c|}
\hline
Source & Events in ROI  & + Topology cuts \\
\hline
$^{232}$Th& 80.5  & 0.5\\
$^{238}$U& 25.8 & 0.7 \\
\hline
Total & 106.3 &  1.2 \\
\hline
\end{tabular}
\caption{Background events per year in the ROI, before and after applying topological selection cuts.}
\label{Tab:radioactivity}
\end{center}
\end{table}%

\subsection{Anode radioactivity}
\label{sec:anode}

The same simulation described in Sec.~\ref{Sec:BGCompo} was used to estimate the background contribution from the anode activity. As explained in Sec.~\ref{Sec:Det}, the use of a hollow tube with a thickness of 1~mm minimises the anode mass to less than 200~g. A statistical sample of 1 million events was required to observe any events passing the selection cuts, corresponding to approximately $1 \times 10^{-3}$ events per year, under the assumption of an activity of 10~$\mu$Bq/kg, as in Sec.~\ref{Sec:BGCompo}. Consequently, the background associated with the anode can be considered negligible, even if the estimated activity is increased by a factor of 100, {\it i.e.}, assuming an activity of 1~mBq/kg.

\subsection{Lead radioactivity}
\label{sec:lead}

The lead shielding used to screen the detector from external gamma radiation 
could also introduce background due to contamination from $\mathrm{^{238}U}$, 
$\mathrm{^{232}Th}$, and their decay chains. The impact of this background was 
evaluated, assuming activities of $12~\mu\mathrm{Bq/kg}$ for uranium and 
$4~\mu\mathrm{Bq/kg}$ for thorium~\cite{Leonard:2007uv}. It should be noted 
that these values represent upper limits constrained by measurement sensitivity, 
and the actual contamination levels may be lower.  

Through the full analysis chain, this background was found to contribute a 
non-negligible $3.5$ events per year for uranium and $1.9$ events per year 
for thorium. These contributions arise primarily from $\mathrm{^{214}Bi}$ 
and $\mathrm{^{208}Tl}$, produced within the first few inner centimetres of 
the lead shielding. However, if the first $\sim 5~\mathrm{cm}$ of lead are excluded, 
the background is significantly reduced to an almost negligible level of 
$0.1$ events per year. Again, this represents an upper limit.  

To mitigate the contribution of the shielding, several options are being 
considered for the final setup. One option is to include a copper layer a 
few centimetres thick. Ultra-low-radioactivity copper, with contamination 
levels below $1~\mu\mathrm{Bq/kg}$, is commercially available from suppliers 
such as Aurubis or Mitsubishi~\cite{MAJORANA:2016lsk}. Alternatively, a 
water-based passive veto could be employed.

In the present study, a contribution of $0.1$ events per year is assumed, with 
the most practical option to achieve this background level still under evaluation.

\subsection{External gammas}
\label{sec:extgamma}
The background due to external gamma radiation is strongly dependent on the detector’s location. In the present study, the assumed gamma spectra and rate are based on measurements performed at the underground laboratory in Modane (LSM)~\cite{NEMO:2002zps}. Following the same strategy as in Ref.~\cite{Meregaglia:2017nhx}, the gamma spectra are divided into three energy ranges: 1-4~MeV, which primarily includes gammas from natural radioactivity; 4-6~MeV, which accounts for gammas emitted by the detector itself; and 6-10~MeV, which comprises gammas resulting from neutron capture in the surrounding materials.

The expected gamma flux in the 1-4~MeV region is 0.12 events cm$^{-2}$ s$^{-1}$. Gammas were generated in the GEANT4 simulation within a sphere of 280~cm radius, homogeneously and isotropically, resulting in a statistic of approximately $3.7 \times 10^{12}$ events per year. The total simulated statistics is $4 \times 10^{13}$ events, corresponding to more than 10~years of data. Out of this total, 1 event remains after cuts, leading to an expected background of 0.1 events per year. This background can, of course, be reduced by increasing shielding if necessary.

Gammas with energies between 4 and 6~MeV are predominantly due to the detector used for the measurement; therefore, the extrapolation from the LSM measurement (considered as an example of a potential underground laboratory location) to the present sensitivity study may be affected by significant uncertainties. Considering the measured flux of $3.8 \times 10^{-6}$ events cm$^{-2}$ s$^{-1}$ and the simulated statistics of $1 \times 10^{10}$ events, generated on the same spherical surface used for the low-energy gammas, this corresponds to a period of over 80 years. Since no events survive the selection cuts, an upper limit can be computed. The upper bound of a 90\% confidence level (C.L.) interval for a Poisson signal mean, when 0 events are observed, is 2.44~\cite{Feldman:1997qc}, leading to an upper limit of approximately 0.03 events per year.

The last energy region studied is between 6 and 10~MeV, corresponding to gammas emitted from nuclear radiative capture in the materials constituting and surrounding the detector. The measured flux at LSM was $3.2 \times 10^{-6}$ events cm$^{-2}$ s$^{-1}$, resulting in an upper limit of 0.02 events per year, considering the simulated statistics of $1 \times 10^{10}$ events and no events passing the topological selection cuts.
All the results are summarized in Tab.~\ref{Tab:gamma}.

\begin{table}[ht]
\begin{center}
\begin{tabular}{|c|c|c|c|}
\hline
Energy & Events in ROI   & + Topology cut \\
\hline
1-4~MeV& 2.9  & 0.1\\
4-6~MeV& 0.03  & < 0.03\\
6-10~MeV& < 0.02 &  < 0.02 \\
\hline
Total & 2.9 & 0.1 \\
\hline
\end{tabular}
\caption{Gamma background per year in the ROI, without and with topological selection cuts.}
\label{Tab:gamma}
\end{center}
\end{table}%

\subsection{External neutrons}
\label{Sec:extNeutron}
Neutrons emitted from radioactive decays or ($\alpha$,n) reactions in the laboratory environment surrounding the detector, as well as neutrons produced by cosmic muon spallation reactions, could be sources of background. Several specific neutron-related backgrounds are considered below.
\begin{itemize}
\item {\it Neutron captures on detector materials}\\ Neutron captures on materials such as Pb, Cu, or Fe produce high-energy gammas in the energy region between 6 and 10~MeV. These gammas have already been considered in the external gamma section ({\it i.e.} Sec.~\ref{sec:extgamma}).
\item {\it Neutrons produced inside the detector by cosmic muons}\\ Neutrons generated by muons inside the detector can be easily tagged, as the muon deposits a significant amount of energy in the active volume. The only potential issue is deadtime (addressed in Sec.~\ref{sec:PileUp}), but this is negligible given the extremely low muon rate at underground laboratories ({\it e.g.} 4~m$^{-2}$ day$^{-1}$ at LSM).
\item {\it Neutron captures on xenon}\\
Neutrons entering the detector can be captured by xenon, resulting in the emission of several gammas with a total energy of a few MeV. The expected neutron flux of $4 \times 10^{-6}$ events cm$^{-2}$ s$^{-1}$ and the energy spectrum between 2 and 6~MeV were obtained from measurements performed at LSM~\cite{NEMO:2002zps}. Similar to the gamma simulation, neutrons were generated in the GEANT4 simulation on a sphere with a radius of 280~cm, homogeneously and with isotropic directions, yielding statistics of about $1.2 \times 10^{8}$ events per year. GEANT4 was also used to estimate neutron captures on the $^{136}$Xe isotope. The simulated statistics of $3 \times 10^{10}$ events, corresponding to approximately 240 years of data taking, a the total number of background events after cuts of 0.02.
\item {\it Spallation neutrons}\\ The energy of spallation neutrons can reach several GeV, and the induced nuclear recoils may produce signals in the ROI that are difficult to reject. The neutron energy spectrum from Ref.~\cite{Mei:2005gm} was used in the simulation, assuming a flux above 10~MeV of $5 \times 10^{-10}$ events cm$^{-2}$ s$^{-1}$, based on an underground depth of 4800~m.w.e., corresponding to that of LSM. A total of $1 \times 10^{6}$ events was simulated (approximately 64 years of data taking),  yielding an expected rate of 0.02 events per year.

\end{itemize}
The neutron related background expectation is summarized in Tab.~\ref{Tab:neutrons}.

\begin{table}[ht]
\begin{center}
\begin{tabular}{|c|c|c|}

\hline
Source & Events in ROI  & + Topology cut \\
\hline
Capture on $^{136}$Xe  & 6.7  &  0.02\\
Spallation neutrons  & 3.7  &  0.02\\
\hline
Total & 10.4 & 0.04 \\
\hline\end{tabular}
\caption{Neutron background per year in the ROI, without and with topological selection cuts.}
\label{Tab:neutrons}
\end{center}
\end{table}%

\subsection{Radon}
\label{sec:RadonBG}
Radon is one of the ultimate sources of background in low-radioactivity experiments. Cryogenic distillation is an effective method for radon removal~\cite{Murra:2022mlr}, and the nEXO collaboration aims to reach a contamination limit of 0.3~$\mu$Bq/kg~\cite{nEXO:2021ujk}. However, maintaining 1~ton of xenon at this low contamination level during data collection requires 1~ton of liquid nitrogen daily. Although radon’s molecular size is similar to that of xenon, complicating its capture using standard molecular filters, ongoing research explores alternative absorbers~\cite{10.1093/ptep/ptad160}.

A sample of $1 \times 10^7$ radon events was simulated homogeneously in the xenon volume, and the full digitization and analysis chain was applied to the Monte Carlo events. A rejection factor of $3 \times 10^{-7}$ was achieved after cuts. Assuming a radon activity level of 5~$\mu$Bq/kg (corresponding to 1.5~mBq/m$^3$), a conservative estimate relative to the contamination levels expected in xenon experiments (0.3~$\mu$Bq/kg for nEXO~\cite{nEXO:2021ujk}, 1~$\mu$Bq/kg for XENONnT~\cite{XENON:2024wpa}, and 3.5~$\mu$Bq/kg for PandaX-4T~\cite{PandaX:2024jjs}), the resulting background amounts to about 0.2 events per year.

\subsection{Cosmogenic background}
\label{Sec:Cosmogenics}
Cosmogenic nuclei produced by muons passing through the xenon volume could contribute to the background, as these nuclei may undergo beta decays, potentially yielding signals in the ROI. A detailed simulation by the EXO-200 collaboration~\cite{EXO-200:2015edf} identified $^{137}$Xe as the primary contributor to signals near the $^{136}$Xe $Q_{\beta\beta}$. The production rate of $^{137}$Xe depends on both the muon flux and their energy. 

Comprehensive studies have been conducted for natural xenon~\cite{DARWIN:2023uje} and enriched $^{136}$Xe~\cite{NEXT:2020qup} at different underground sites. Assuming a detector filled with enriched $^{136}$Xe, which results in higher $^{137}$Xe production, the expected rate at LSM is conservatively estimated to be $1.3 \times 10^{-2}$ kg$^{-1}$ year$^{-1}$. This value was derived from studies at SURF (4300 m.w.e.), which has a lower depth than LSM (4800 m.w.e.), and corresponds to an expected production rate of approximately 10 events per year in the proposed setup. Given a selection efficiency of $1.6 \times 10^{-4}$, determined from the full simulation chain, this background source can be considered negligible.

A muon veto does not seems necessary given the weak production rate of $^{137}$Xe, however it could be considered if needed depending on the actual experiment location.

\subsection{Background summary}

\begin{figure}[tp]
     \centering     
   	 \subfigure[\label{fig:BGSpectrumBS}]{\includegraphics[width=\columnwidth]{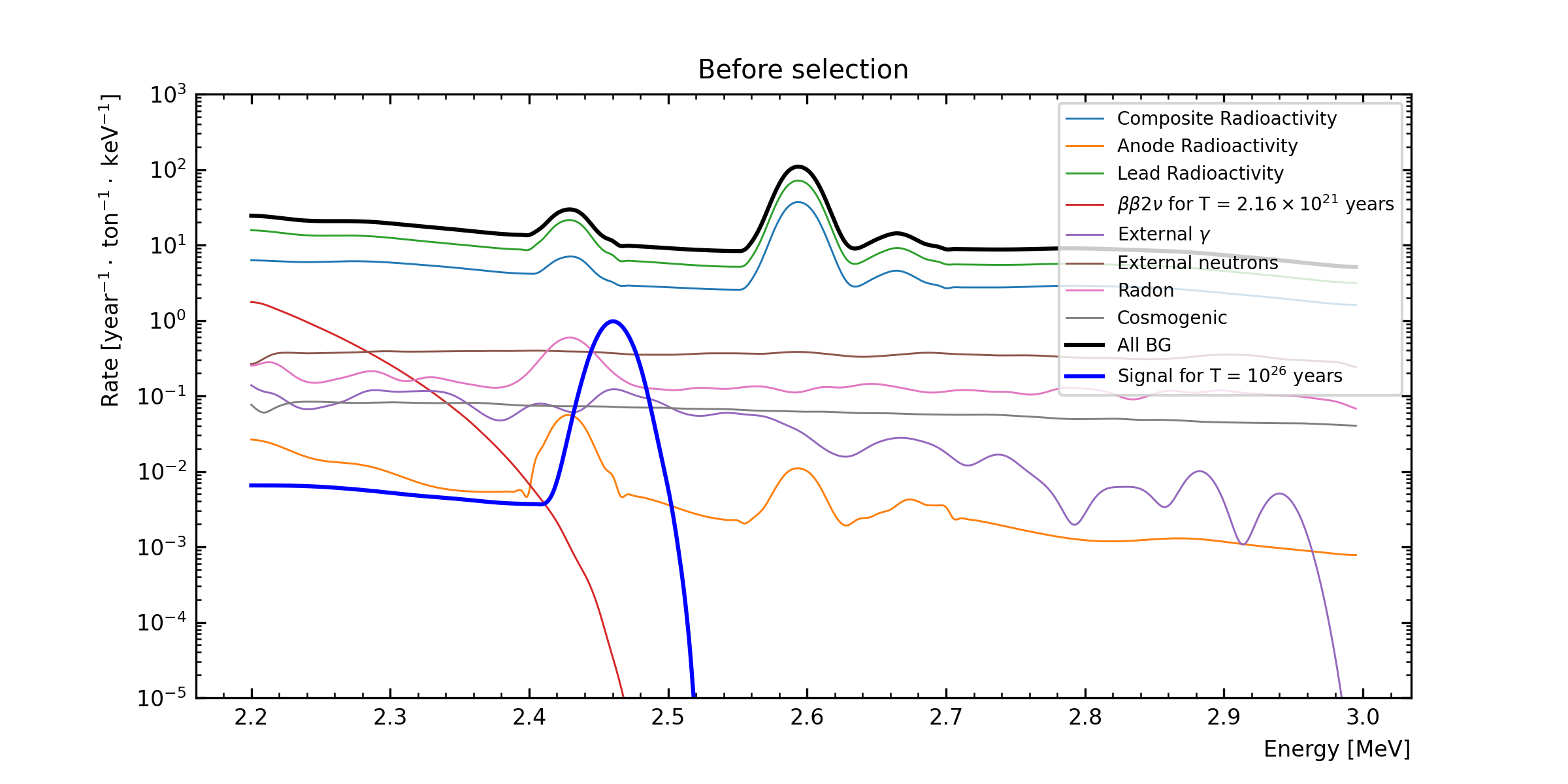}}
      \subfigure[\label{fig:BGSpectrumAS}]{\includegraphics[width=\columnwidth]{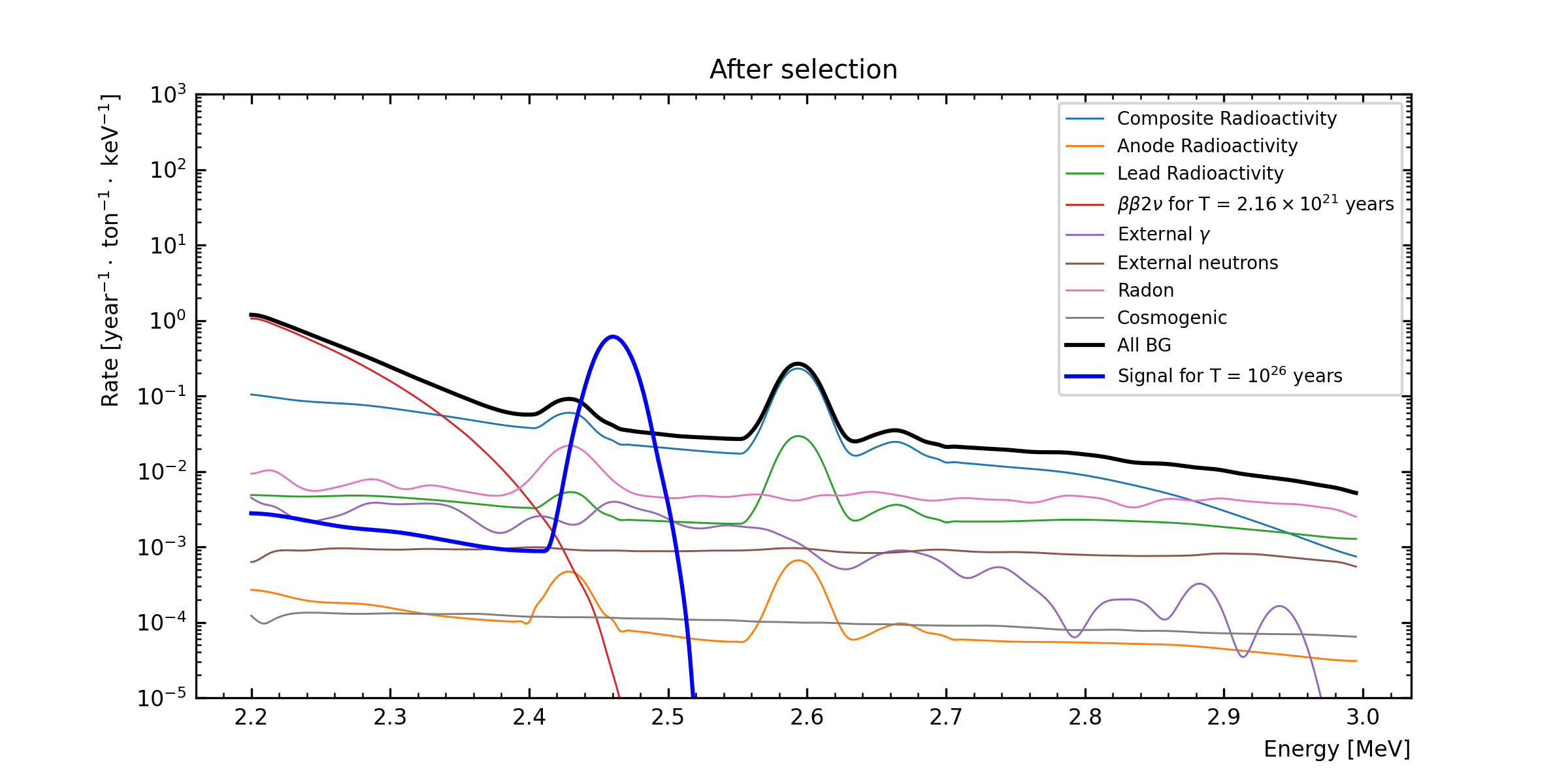}}
    \caption{Signal and background energy spectrum before~\ref{fig:BGSpectrumBS} and after ~\ref{fig:BGSpectrumAS} selection cuts. The different components (thin lines) as well the sum of all the background (thick black line) are shown. For comparison a signal expected for a mean lifetime of $T^{0\nu}_{1/2} = 10^{26}$ years is also shown (thick blue line).}
    \label{fig:BGSpectrum}
\end{figure}

Various potential sources of background were considered for the proposed detector. Assuming the CPC is filled with pure $^{136}$Xe and situated underground at 4800 m.w.e. (the depth of LSM), the expected background level is approximately 1.6 events per year. This estimate may vary slightly depending on the specific underground laboratory, influenced by differences in overburden and environmental radioactivity. However, the result is valid for similar sites, such as LNGS in Italy. A summary of the background contributions after selection cuts is provided in Tab.~\ref{Tab:BGSummary}, where a 90\% confidence level upper limit is used in cases where no events passed the cuts in the simulation.

The background energy spectrum is also shown in Fig.~\ref{fig:BGSpectrum} before and after the application of the selection cuts.

\begin{table}[ht]
\begin{center}
\begin{tabular}{|c|c|c|}
\hline
Contribution & Events per year\\
\hline
$\beta\beta2\nu$ & $1.0 \times 10^{-3}$ \\
Composite radioactivity & 1.2 \\
Anode radioactivity & $1.0 \times 10^{-3}$ \\
Lead radioactivity & 0.1  \\
External gammas & 0.1 \\
External neutrons & 0.04 \\
Radon & 0.2 \\
Cosmogenic background & $1.6 \times 10^{-3}$ \\
\hline
Total &  1.6\\
\hline
\end{tabular}
\caption{Summary of background contributions after selection cuts.}
\label{Tab:BGSummary}
\end{center}
\end{table}%

\section{Pile Up}
\label{sec:PileUp}

The total event rate is a critical parameter for the experiment. Given the long drift time, of the order of a few milliseconds, even events outside the ROI, though not directly contributing to the background, could overlap with good events, compromising their energy or position reconstruction. To maintain a minimal pile-up probability of $5 \times 10^{-3}$, assuming a time window of 5~ms which corresponds  to the maximum expected drift time, the event rate should be kept below 1~Hz. The same background sources considered in Sec.~\ref{Sec:BG} were re-evaluated with a significantly lower energy threshold, down to a few keV.

As computed in Sec.~\ref{Sec:BGbb}, the number of $\beta\beta2\nu$ events expected per year is approximately 825000, corresponding to an event rate of $\sim 0.03$~Hz.

The radioactivity of the composite material was simulated in Sec.~\ref{Sec:BGCompo}, focusing on $^{232}$Th and $^{238}$U, with an assumed activity of 10~$\mu$Bq/kg. The fraction of events releasing energy within the xenon active volume is 8\% and 5\%, respectively. Based on the expected number of events, the rate for each of these decay chains is $2.2\times 10^{-3}$~Hz, yielding a total rate of $4.4\times 10^{-3}$~Hz.

The activity of the anode has a negligible impact on pile-up, as its small mass results in an expected event rate of approximately 1500 events per year, corresponding to $4.7 \times 10^{-5}$~Hz.

The activity of lead also has a negligible impact on pile-up, since even without 
any reduction, as discussed in Sec.~\ref{sec:lead}, the contribution to the 
pile-up is at the level of $8.0 \times 10^{-3}$~Hz.

The external gamma simulation from Sec.~\ref{sec:extgamma} was used to calculate the fraction of events depositing energy in the active volume. Out of $1 \times 10^{8}$ simulated events between 1 and 4~MeV, only one event deposited energy in the xenon active volume. Scaling this fraction to the expected number of events per year yields an event rate of $1.2\times 10^{-3}$~Hz. For the 4 to 6~MeV and 6 to 10~MeV ranges, no events deposited energy in the active volume within the simulated $1 \times 10^{7}$ events. This results in an upper limit of approximately 29 and 24 events per year, respectively, corresponding to a negligible rate of $\sim 1\times 10^{-6}$~Hz in both cases.

The external neutrons were discussed in Sec.~\ref{Sec:extNeutron}. A total of $1 \times 10^{6}$ events were generated, with a fraction of $\sim 1.4 \times 10^{-4}$ producing a signal inside the detector. Based on the expected neutron flux per year, this contribution results in a trigger rate of approximately $5 \times 10^{-4}$~Hz. The contribution from spallation neutrons is significantly smaller, given that the expected flux is lower by four orders of magnitude. Despite a fraction of events inducing a signal at the level of $5 \times 10^{-2}$, the resulting rate is $2.5 \times 10^{-5}$~Hz.

Radon background was discussed in Sec.~\ref{sec:RadonBG}, where an activity level of 5~$\mu$Bq/kg was assumed. With 76\% of events producing a signal in the active volume, this corresponds to approximately 617000 events per year, resulting in a rate of 0.02~Hz.

Finally, the cosmogenic background discussed in Sec.~\ref{Sec:Cosmogenics} must be considered. For pile-up studies, the relevant factor is the number of muons crossing the detector rather than the isotopes produced. Using the expected muon rate at LSM of 4~m$^{-2}$ day$^{-1}$ and conservatively assuming a circular surface above the detector with a 10~m radius (i.e., muons with zenith angles up to 80~degrees), approximately 1256 muons per day are expected. If all of these muons are assumed to pass through the detector, the resulting rate is 0.015~Hz.

Additionally, an extra source of pile-up could arise from $^{40}$K contamination in the composite material. Although this source was not studied in the background section, as the maximum energy released in the decay (1.5~MeV) is outside the ROI, $^{40}$K contamination in glass is a known issue in experiments using photomultiplier tubes for signal detection. It should therefore also be considered in composite materials. The fraction of simulated events depositing energy in the xenon volume is 3.7\%. No direct measurements have yet been performed on the composite materials proposed for the tank, so the contamination level was estimated based on glass measurements~\cite{Cuesta:2014zga}. Assuming an activity of 1~mBq/kg, the expected rate is 0.12~Hz, representing the largest contribution to the pile-up.

The total event rate expected in the detector is approximately 0.20~Hz (see Tab.~\ref{Tab:PileUpSummary}), which is significantly below the acceptable limit of 1~Hz, considered necessary for safe detector operation.

\begin{table}[ht]
\begin{center}
\begin{tabular}{|c|c|c|}
\hline
Contribution & Rate (Hz)\\
\hline
$\beta\beta2\nu$ & 0.03 \\
Composite radioactivity & $4.4\times 10^{-3}$\\
Anode radioactivity & $4.7 \times 10^{-5}$ \\
Lead radioactivity & $8.0 \times 10^{-3}$ \\
External gammas & $1.2\times 10^{-3}$ \\
External neutrons & $5 \times 10^{-4}$ \\
Radon & 0.02 \\
Cosmogenic background & 0.015 \\
$^{40}$K & 0.12 \\
\hline
Total & 0.20\\
\hline
\end{tabular}
\caption{Expected event rates in the detector for various contributions.}
\label{Tab:PileUpSummary}
\end{center}
\end{table}%

\section{Sensitivity results}
\label{Sec:sensitivity}
\begin{figure}[t]
    \centering
    \includegraphics[width=\columnwidth]{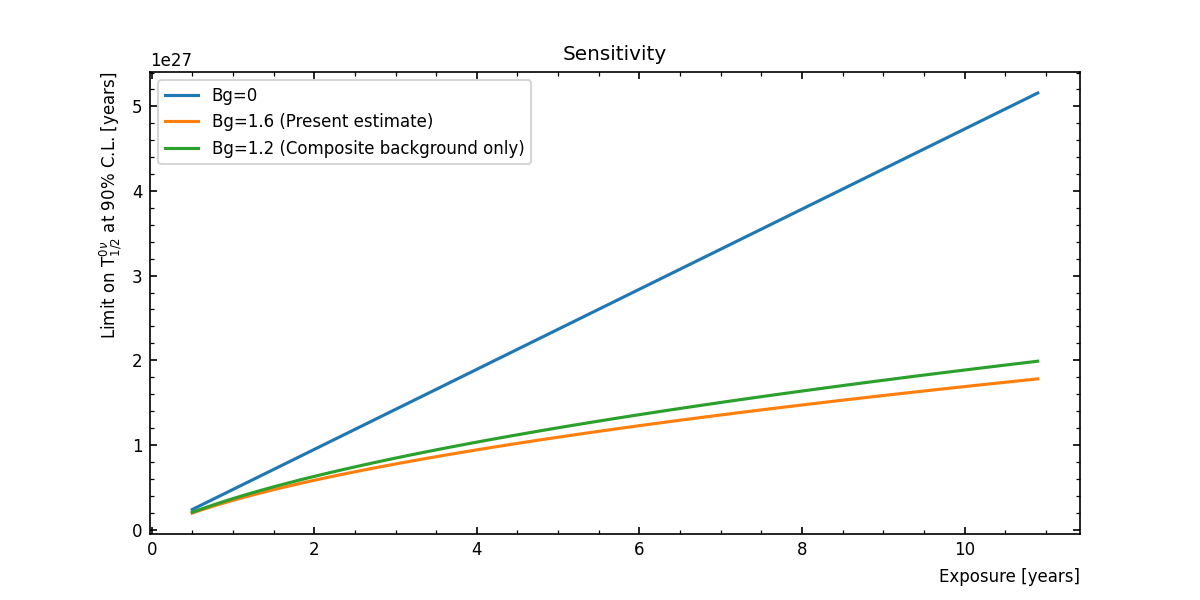}
  \caption{Expected sensitivity at 90\% C.L. for zero background (blue line), for the expected background of 1.6 events per year (orange line), and for the contribution due to composite radioactivity only (green line).}
    \label{fig:Sensitivity1}
\end{figure}

The experimental sensitivity can be computed in terms of a limit on the half-life, assuming complete enrichment of $^{136}$Xe. The limit can be expressed as:
\begin{equation}
\label{eq:nobg}
T^{0\nu}_{1/2} > \ln(2) \epsilon \frac{N_A m}{M} \frac{t}{S_{up}},
\end{equation}
where $\epsilon$ is the signal efficiency, $N_A$ is Avogadro’s number, $M$ is the molar mass of xenon in grams, $t$ is the exposure in years, and $m$ is the active mass of xenon in grams. The signal upper limit, $S_{up}$, can be computed using a Bayesian upper limit for a Poisson parameter with a uniform prior probability density function (p.d.f.), as explained in Ref.~\cite{ParticleDataGroup:2022pth}:
\begin{equation}
S_{up} = \frac{1}{2} F^{-1}_{\chi^2} [p,2(n+1)] -b,
\end{equation}
where $F^{-1}_{\chi^2}$ is the quantile of the $\chi^2$ distribution ({\it i.e.} the inverse of the cumulative distribution), $b$ the expected background, and $n$ the number of observed events. The quantity $p$ is defined as:
\begin{equation}
p=1 - \alpha(1 - F_{\chi^2}[2b,2(n+1)]),
\end{equation}
where $F_{\chi^2}$ is the cumulative $\chi^2$ distribution, and (1-$\alpha$) is the confidence level.

This sensitivity is applicable regardless of the experimental background; however, in the case of a background-free experiment, $S_{up}$ is constant, and the limits depend linearly on the exposure. This is not true for an experiment with background present: the background also increases with exposure (and therefore $S_{up}$), resulting in a sensitivity dependence that is proportional to the square root of the exposure under the Gaussian description of background fluctuations. The results for different background values can be seen in Fig.~\ref{fig:Sensitivity1}. With the nominal expected background, a sensitivity on $T^{0\nu}_{1/2}$ of $1.1 \times 10^{27}$ years can be achieved in 5 years of data taking ($1.7 \times 10^{27}$ years in 10 years of data taking). It is clear that the sensitivity is dominated by the background from the composite material, and to improve it, efforts must be undertaken to develop a thinner vessel that withstands the operating pressure of 40 bars, or one with reduced radioactivity.

\begin{figure}[t]
    \centering
    \includegraphics[width=\columnwidth]{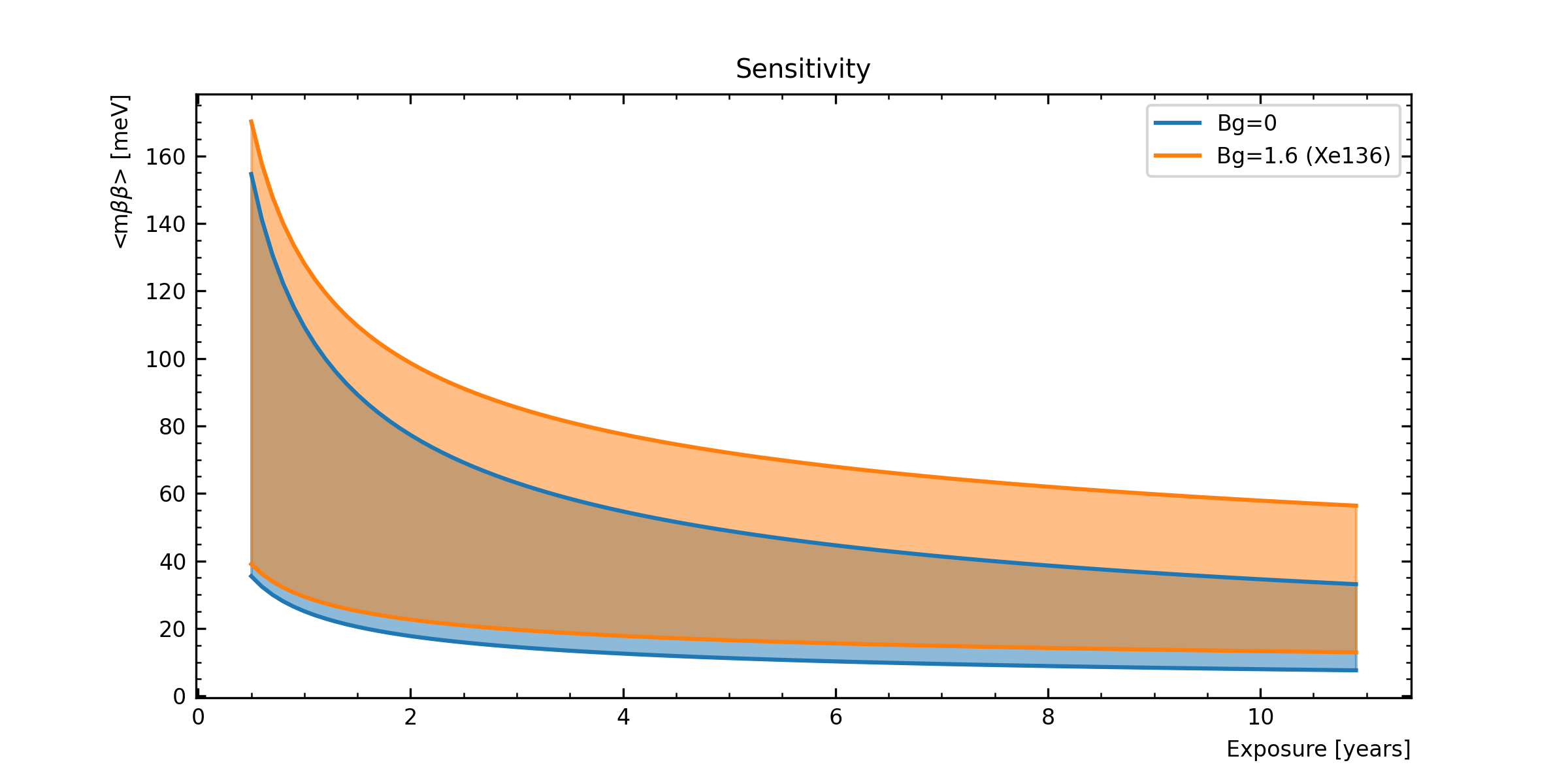}
    \caption{Expected limits for $<m_{\beta\beta}>$ at 90\% C.L. for zero background (blue region) and for an expected background of 1.6 events per year (orange band). The width of the band represents the uncertainty in the nuclear matrix elements.}
    \label{fig:Sensitivity2}
\end{figure}

The lifetime is related to the effective Majorana mass $<m_{\beta\beta}>$ by the following formula:
\begin{equation}
\frac{1}{T^{0\nu}_{1/2}}=G^{0\nu} |M^{0\nu}|^2 <m_{\beta\beta}>^2,
\end{equation}
where $G^{0\nu}$ represents the phase space factor and $M^{0\nu}$ is the matrix element. Using the values from Ref.~\cite{KamLAND-Zen:2022tow} and the references therein, the limit on $<m_{\beta\beta}>$ was computed both for the zero-background scenario and for the current background estimate, with the results shown in Fig.~\ref{fig:Sensitivity2}. Considering the expected background, the limit on $<m_{\beta\beta}>$ lies in the range of 13~meV to 57~meV after 10 years, which is reduced to 8~meV to 35~meV if zero background is achieved.

\section{R2D2 in the worldwide panorama}
The present best senstivities on the effective neutrino mass have been obtained by KamLAND-Zen~\cite{KamLAND-Zen:2024eml} and GERDA~\cite{GERDA:2020xhi}. KamLAND-Zen measured $2.097~\mathrm{ton \cdot yr}$ of $^{136}$Xe, achieving a $\beta\beta0\nu$ half-life limit of $T_{1/2}^{0\nu} > 3.8 \times 10^{26}~\mathrm{years}$ 
at 90\% C.L., corresponding to an effective neutrino mass $<m_{\beta\beta}>$ between 28 and 122~meV. GERDA measured 
$125~\mathrm{kg \cdot yr}$ of $^{76}$Ge, setting a half-life limit of 
$T_{1/2}^{0\nu} > 1.8 \times 10^{26}~\mathrm{years}$, which corresponds to 
$<m_{\beta\beta}>$ between 79 and 180~meV.

The performance of the proposed experiment depends on the final background and could be compared with future multi-ton experiments. In particular, the following experiments were considered: nEXO~\cite{nEXO:2021ujk}, LEGEND1000~\cite{LEGEND:2021bnm,LEGEND1000}, and CUPID~\cite{CUPID:2022wpt}. 

\begin{figure}[t]
    \centering
    \includegraphics[width=\columnwidth]{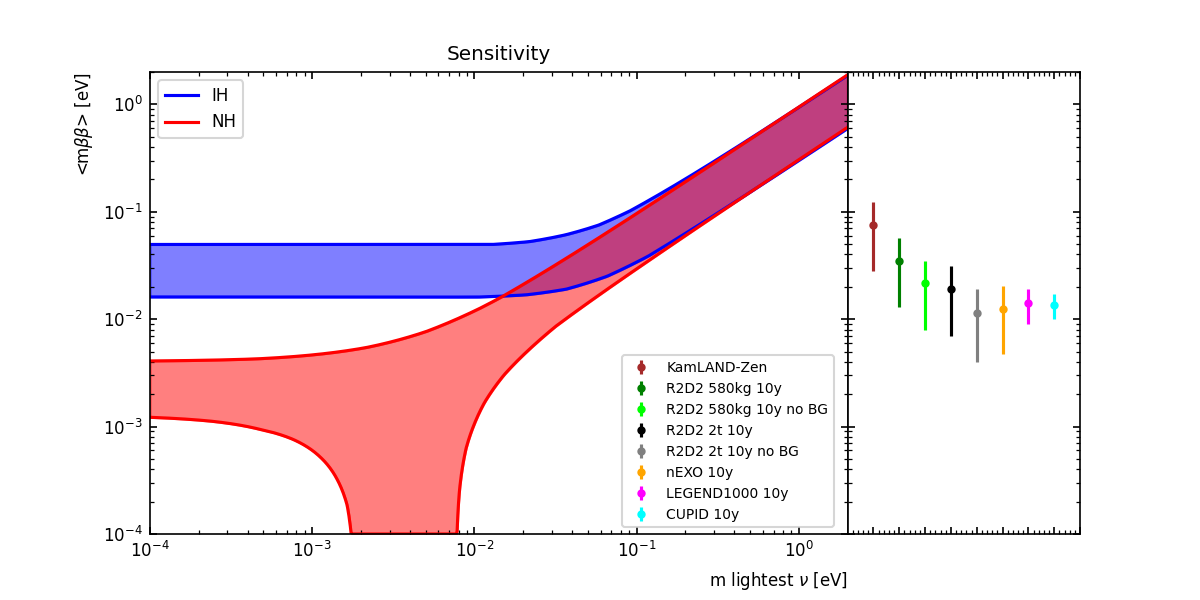}
    \caption{Current and expected limits for $<m_{\beta\beta}>$ at 90\% C.L. for various experiments.}
    \label{fig:SensitivityComparison}
\end{figure}

As shown in Fig.~\ref{fig:SensitivityComparison}, the performance of R2D2 is only a factor of 2.5 worse with respect to that of future proposed experiments and could be comparable if zero background is achieved.
A reduction of the external gamma component (0.1 events per year) is relatively straightforward, as it depends solely on shielding, whereas the ultimate background arises from composite radioactivity (1.2 events per year) and radon contamination (0.2 events per year). As previously explained, a radioactivity level of 10~$\mu$Bq/kg is considered in the present study, but efforts are ongoing to potentially reduce this value. Furthermore, ongoing studies indicate a possible reduction in the thickness of the composite vessel by a factor of 6, resulting in a total vessel mass of only about 60~kg.
Multiple detectors could also be built, increasing the active mass. The projected sensitivity for a 2~ton detector is also shown in Fig.~\ref{fig:SensitivityComparison} for comparison.

Another point of comparison could be costs and timescale. The timescale is nearly identical for all next-generation experiments, which aim to yield final results around 2040. In this context, R2D2 could be timely, as final results could be expected around 2040 if the program is fully funded and data collection commences in 2030. The cost of the detector itself is low ($\sim 3$~M\texteuro), excluding the cost of enriched xenon. Moreover, no cryogenic infrastructure is required, and the power consumption is minimal, which helps reduce the environmental impact.

Last but not least, given their cost, this type of detector could be deployed in pairs: one filled with natural xenon and the other with $^{136}$Xe. Another advantage could lie in the simultaneous deployment of these duplicates in different underground laboratories. This strategy would allow for differential measurements of signal and noise, providing the opportunity to gain a more comprehensive understanding of experimental observations. Coupled with the ability to vary the nature of the medium ({\it i.e.} Ar or Xe in the adjustment phase), the isotopic content of the gas (natural Xe or $^{136}$Xe in the working phase), or to modulate the pressure (up to 50~bars), the proposed scenario could offer valuable redundancies to verify any potential evidence of signals.

\begin{acknowledgements}
The authors would like to express their gratitude for the support received from various funding sources that contributed to this research. The R\&D funding from IN2P3 provided essential resources for the development of the project. Additionally, the OWEN grant from Bordeaux University contributed to the development of low noise electronics. Grants from SUBATECH significantly contributed to the advancement of the research objectives. The authors also acknowledge the use of the computing cluster at the CC Lyon of IN2P3, which was invaluable for the computational aspects of this work.
\end{acknowledgements}

\bibliographystyle{spphys}       
\bibliography{R2D2bibliography}   

\end{document}